\documentclass[aps,pra,twocolumn,amsmath,amssymb,nofootinbib,superscriptaddress]{revtex4-1}

\newcommand{\bra}[1]{\left\langle#1\right\rvert}
\newcommand{\ket}[1]{\left\lvert#1\right\rangle}
\newcommand{\ketbra}[2]{\left\lvert{#1}\middle\rangle\!\middle\langle{#2}\right\rvert}

\newcommand{\expect}[1]{\left\langle{#1}\right\rangle}

\newcommand{\tr}{\mathrm{Tr}}

\usepackage[colorlinks]{hyperref} 
\usepackage[pdftex]{graphicx}
\usepackage{subfigure}
\usepackage{mathrsfs}
\usepackage[usenames,dvipsnames]{xcolor}
\usepackage{framed}

\definecolor{cream}{rgb}{1.0, 0.99, 0.82}
\definecolor{celadon}{rgb}{0.67, 0.88, 0.69}
\definecolor{beaublue}{rgb}{0.74, 0.83, 0.9}
\definecolor{bole}{rgb}{0.47, 0.27, 0.23}
\definecolor{shadecolor}{rgb}{1.0, 0.99, 0.82}

\usepackage{soul}

\hypersetup{
    colorlinks=true, 
    linktoc=all,     
    linkcolor=blue,  
    citecolor=blue,
    filecolor=blue,
    urlcolor=blue
}

\begin{document}

\title{Power and Efficiency of a Thermal Engine with a Coherent Bath}

\author{Thomas Guff}
\email[]{thomas.guff@students.mq.edu.au}
\affiliation{Centre for Engineered Quantum Systems, Department of Physics and Astronomy, Macquarie University, Sydney NSW 2113, Australia}

\author{Shakib Daryanoosh}
\affiliation{Centre for Engineered Quantum Systems, Department of Physics and Astronomy, Macquarie University, Sydney NSW 2113, Australia}

\author{Ben Q. Baragiola}
\affiliation{Centre for Engineered Quantum Systems, Department of Physics and Astronomy, Macquarie University, Sydney NSW 2113, Australia}

\author{Alexei Gilchrist}
\affiliation{Centre for Engineered Quantum Systems, Department of Physics and Astronomy, Macquarie University, Sydney NSW 2113, Australia}

\date{\today}

\begin{abstract}
We consider a quantum engine driven by repeated weak interactions with a heat bath of identical three-level atoms. This model was first introduced by Scully et al. [Science, 2003], who showed that coherence between the energy-degenerate ground states serves as a thermodynamic resource that allows operation of a thermal cycle with a coherence-dependent thermalisation temperature. We consider a similar engine out of the quasistatic limit and find that the ground-state coherence also determines the rate of thermalisation, therefore increasing the output power and the engine efficiency only when the thermalisation temperature is reduced; revealing a more nuanced perspective of coherence as a resource. This allows us to optimise the output power by adjusting the coherence and relative stroke durations. 
\end{abstract}

\maketitle

\section{Introduction} 
Quantum thermodynamics is concerned with how manifestly qfuantum phenomena, such as coherence and entanglement, affect thermal processes. A primary objective \cite{2016Goold143001,2016Vinjanampathy545} is understanding how these quantum phenomena can be harnessed to improve thermal processes, for example to increase work extraction from a heat engine. Conversely, what limitations does quantum mechanics place on thermal processes? 
There has been research investigating the work extracted from a quantum measurement \cite{Erez2012,Kammerlander2016,Jacobs2012,Ding2018,Yi2017}, or measurements with feedback \citep{Funo2013,Maruyama2009,Elouard2017,Elouard2018}, known as a Maxwell's demon. Systems initialised in non-equilibrium states allow for more work extraction according to generalisations of the second law \cite{Hasegawa2010,Niedenzu2018}. Correlations between systems can also increase the extractable work \cite{Oppenheim2002,Perarnau-Llobet2015,Zurek2003,Jevtic2012,Rio2011}, and there is a thermodynamic cost to creating correlations in uncorrelated, thermal systems \cite{Esposito2010}.

Of particular interest is the effect of coherence in thermal processes. Although it contributes to the free energy, coherence cannot be extracted as work using thermal operations \cite{Lostaglio2015,Korzekwa2016}, but can be used as a catalyst \cite{Aberg2014} and thought of as a quantum resource \cite{Baumgratz2014,Lostaglio2015X,Marvian2016}.
Instead of considering coherence in the system itself Scully et al. \cite{2003scully862} studied a system which interacted with a coherent heat bath. Specifically they considered a photo-Carnot engine ---  a single mode cavity undergoing a Carnot cycle --- in which a heat bath is emulated by repeated weak interactions with identical three-level atoms with two degenerate ground states. Coherence between the two ground states can change the thermalisation temperature of the cavity, so a Carnot cycle can be implemented using a single bath and switching on coherence. There has been extensions to this setup, including adding cavity leakage \cite{Quan2006} and using entangled qubits \cite{Dillenschneider2009}, or more general multi-atom baths \cite{Turkpence2017,Dag2016,Dag2018,Hardal2015}.

In the quasistatic limit, the efficiency is maximal, but since this requires infinite cycle duration, the output power is zero and the engine is impractical; one must trade off engine efficiency for output power \cite{Curzon1975}. We investigate the effect on the power and efficiency of a quantum harmonic oscillator undergoing an Otto cycle with a coherent heat bath. Following Scully et al. we emulate a heat bath using repeated weak interactions with identical three-level systems with thermal populations and coherence between the two degenerate energy ground states. The interaction between the three-level atoms and the system is governed by a Jaynes-Cummings Hamiltonian, from which we derive a master equation for the state of the system.  This allows us to analyse the performance of the engine in a more practical regime; without requiring the duration of each stroke to be infinite.

The two isentropic strokes involve changing the harmonic frequency $\omega$. If this occurs rapidly then many off-diagonal coherences will be excited in the system. To correct for this effect we use shortcut-to-adiabaticity techniques (see \cite{ATOMTorrontegui2013} for a review). These techniques allow the state to undergo effectively adiabatic evolution in a finite time, and can be used to drive a thermal engine \cite{Campo2015,Beau2016,Abah2018,Abah2018PRE}. In \cite{EPLLutz2017}, the authors calculate a cost which must be paid to perform their shortcut-to-adiabaticity protocol, which increases with shorter stroke times. We adopt this cost as the reduction in work extracted during the Otto cycle.

This paper has the following structure. In section~\ref{sec:interaction} we describe the interaction between the bath and the system, showing the change in the thermalisation temperature of the system when there is coherence between the ground states of the atoms. In section~\ref{sec:Otto} we describe the four strokes of the Otto cycle, and in section~\ref{sec:finite} we describe some issues with running an engine in finite time. Section~\ref{sec:results} contains our results of the effect of coherence on the power and efficiency of the heat engine in the finite time regime; and in section~\ref{sec:optimise} we optimise the output power of the heat engine using coherence and by changing the relative duration of each stroke.

\section{Interaction with Bath}\label{sec:interaction}

The working fluid of the heat engine is a quantum harmonic oscillator with unit mass, described by the Hamiltonian
\begin{equation}
H_{S} = \frac{1}{2}\left(p^{2}+\omega^{2}x^{2}\right) = \hbar \omega \left( a^\dagger a + \frac{1}{2}\right).
\end{equation}
where $a^{\dagger}a$ is the usual photon number operator. 
A sequence of $(\ell+1)$-level atoms interact for a short time with the system, and after many weak interactions this stream of atoms emulates a 
thermal bath as it causes the system to thermalise. Here we leave the number of atomic energy levels general, but we will shortly assume that $\ell=2$. The atoms have $\ell$ degenerate ground states $\ket{g_j}$ and a single excited 
state $\ket{e}$ with an energy level spacing tuned to the system frequency $\omega$,
\begin{equation}
H_{R} = \hbar \omega \ketbra{e}{e}.
\end{equation}
In this paper we consider loss to the environment to be negligible and focus on the interaction with the atoms as the dominant process. 

The interaction between a single atom and the system is described by the familiar Hamiltonian in the rotating wave approximation
\begin{equation}\label{eq:Hi}
  H_{I} =  \hbar \Omega \left(\sum_{j=1}^\ell a^\dagger  \ketbra{g_j}{e} + a  \ketbra{e}{g_j}\right),
\end{equation}
where $\Omega$ is the interaction strength. Introducing \mbox{$J_-=\ketbra{G}{e}$} and $J_+=J_-^\dagger$, where \mbox{$\ket{G}=\sum_j  \ket{g_j}/\sqrt{\ell}$},
we can write the interaction Hamiltonian as 
\mbox{$H_\mathrm{I} =  \hbar \sqrt{\ell}\Omega (a^\dagger  J_-+ a  J_+)$}. In the interaction picture with zero detuning, the total Hamiltonian reduces simply to the interaction Hamiltonian $H_{I}$. Thus the interaction with duration $\tau$ is governed by the unitary 
\begin{align*}
  U &= e^{-i H_{\mathrm{I}} \tau/\hbar}=e^{-i\lambda\tau(a^\dagger  J_-+ a  J_+)} \\
  &\approx I - i\lambda\tau (a^\dagger  J_-+ a  J_+) -\frac{\lambda^2\tau^{2}}{2}(a^\dagger a J_-J_+ + aa^\dagger J_+J_-)
\end{align*}
where we have expanded to second order in $\lambda\tau=\sqrt{\ell}\Omega \tau$ which is assumed to be small.

Assuming the atom starts in the initial state $\rho_R$ at some time $t'$, each interaction induces a completely positive trace preserving (CPTP) map on the system-atom product state $\rho\otimes \rho_{R}$:
\begin{align}
	&\rho\left(t^{\prime}+\tau\right) = \mathcal{E}(\rho\left(t^{\prime}\right)) =  \mathrm{Tr}_R \{U (\rho\left(t^{\prime}\right)\otimes\rho_R) U^\dagger\} \label{eq:cptp-map}\\
	&\approx \rho\left(t^{\prime}\right) + \lambda^2 \tau^{2} (\bra{e}\rho_R\ket{e} \mathcal{D}[a^\dagger] +
	\bra{G}\rho_R\ket{G} \mathcal{D}[a])\rho\left(t^{\prime}\right), \nonumber
\end{align}
where $\mathcal{D}[b]\rho = b\rho b^\dagger -1/2 b^\dagger b \rho -1/2 \rho b^\dagger b$ is the Lindblad superoperator. In the derivation of \eqref{eq:cptp-map} we assumed that the atoms were block-diagonal in the energy basis, consequently
\begin{align}\label{eq:meanzero}
    \bra{e}\rho_{R}\ket{G}=0,
\end{align}
which removes the term of order $\lambda\tau$. It is straightforward to show that this map contains thermal states as fixed points.

To derive a master equation, we first note that since $\lambda^{2}\tau^{2}$ is assumed small, the map \eqref{eq:cptp-map} is equivalent to $N$ applications of a weaker imaginary map of time duration $\tau_N = \tau/N$:
\begin{multline*}
\rho\left(t^{\prime}+\tau\right) = \\
\left(I +\lambda^2 \tau\tau_N \left(\bra{e}\rho_R\ket{e} \mathcal{D}[a^\dagger] + \bra{G}\rho_R\ket{G} \mathcal{D}[a]\right)\right)^{N}\rho\left(t^{\prime}\right).
\end{multline*}
Considering a single step of this weaker map we can go to the continuum limit 
\begin{equation}
\frac{d\rho(t^{\prime})}{dt^{\prime}} = \lim_{\tau_N\rightarrow 0} \frac{\rho\left(t^{\prime}+\tau_N\right)-\rho\left(t^{\prime}\right)}{\tau_N},
\end{equation}
which yields a dissipative Lindblad master equation
\begin{equation}
	\frac{d\rho(t)}{dt} = \bra{e}\rho_R\ket{e} \mathcal{D}[a^\dagger] \rho(t) +
	\bra{G}\rho_R\ket{G} \mathcal{D}[a] \rho(t), \label{eq:me}
\end{equation}
where for simplification we have rescaled time to the dimensionless parameter \mbox{$t=\lambda^2 \tau t^{\prime}$}.

If the initial state of the system is diagonal in photon number, the steady state of the master equation satisfies
\begin{align}
	\bra{n}\rho\ket{n} = \frac{(E/G)^n}{\sum_n (E/G)^n} = \frac{G-E}{G}  \left( \frac{E}{G} \right)^n ,
\end{align}
where the labeling $E=\bra{e}\rho_R\ket{e}$ and $G=\bra{G}\rho_R\ket{G}$ has been introduced to simplify the expressions.

If the atoms are prepared in a thermal state at inverse temperature $\beta_R$ then
\begin{align}
\rho_R = p_e \ketbra{e}{e} + \frac{1-p_e}{\ell}\sum_j \ketbra{g_j}{g_j},
\end{align}
where
\begin{equation}
p_e = \exp(-\hbar \omega \beta_R)/Z, \label{eq:pe}
\end{equation}
and the partition function is $Z = \ell+\exp(-\hbar \omega \beta_R)$. In this case $E=p_e$ and $G=(1-p_e)/\ell = 1/Z$, and the steady state of the master equation~\eqref{eq:me} is a thermal distribution at inverse temperature $\beta = \beta_R$,
\begin{align}
\rho \rightarrow \gamma_\beta = \frac{e^{-\beta \hbar a^\dagger a}}{\tr(e^{-\beta \hbar a^\dagger a})}.
\end{align}
So the atoms, despite being a rather structured interaction, are emulating a thermal bath at a fixed temperature to which the system thermalises. It can be shown that if the system begins in a thermal state, then it remains in a thermal state under the evolution of this master equation, equilibrating to a particular temperature in the steady state. 

In a thermodynamic analysis we are less interested in the explicit density matrix, than the internal energy. Changes in average energy are proportional to changes in average photon number, $\Delta \bar{H}_{S} = \hbar\omega \Delta \bar{n}$.
Master equation~(\ref{eq:me}) allows us to derive a time evolution equation for the mean photon number $\bar{n}=\tr(a^\dagger a \rho)$,
\begin{align}
	\frac{d\bar{n}}{dt} = E(\bar{n}+1)-G\bar{n}. \label{eq:dndt}
\end{align}
This equation has the solution
\begin{align}
	\bar{n}(t) = \left( \bar{n}(0) -\frac{E}{G-E}\right)e^{-(G-E)t}+ \frac{E}{G-E}. \label{eq:nt}
\end{align}
The steady state limit exists provided $G > E$ and is 
\begin{align}
\bar{n}_\mathrm{ss} = \lim_{t\rightarrow \infty} \bar{n}\left(t\right) = \frac{E}{G-E}.
\end{align}
If $G\leq E$ then the average photon number will increase unbounded and the system will not thermalise. While this can be achieved using coherence, the atoms cease to behave similarly to a thermal bath and so we exclude this case.\\
For simplicity, from here onwards we will use \mbox{$\Delta \equiv G-E$}.

In the steady state limit, the harmonic oscillator is in a thermal state, and the steady state average photon number is related to the inverse temperature as
\begin{equation}
\bar{n}_{\mathrm{ss}} = \frac{1}{e^{\beta\hbar\omega}-1},
\end{equation}
and therefore system will thermalise to the temperature
\begin{equation}
	\beta = \frac{1}{\hbar\omega}\ln\left(\frac{\bar{n}_{\mathrm{ss}}+1}{\bar{n}_{\mathrm{ss}}}\right) = \frac{1}{\hbar\omega}\ln\left(\frac{G}{E}\right).
\end{equation}
The temperature is  strictly monotonic in $\bar{n}_{\mathrm{ss}}$, so we can consider $\bar{n}_{\mathrm{ss}}=E/(G-E)$ as the effective temperature of the atoms. Hence in section~\ref{sec:results} when we compare different pairs of thermal baths with the same effective temperatures, we are holding $\bar{n}_{\mathrm{ss}}$ constant.

The value of $G=\bra{G}\rho_{R}\ket{G}$ (and hence $\bar{n}_{\mathrm{ss}}$ and $\beta$ too) is altered by coherence in the energy ground space of the atoms. In general the atoms are block-diagonal in energy
\begin{equation}
\rho_{R} = p_{e}\ketbra{e}{e} + (1-p_{e})\rho_{g},
\end{equation}
where $p_{e}$ is defined as it was previously \eqref{eq:pe}, and $\rho_{g}$ is the quantum state in the degenerate ground subspace. Hence \mbox{$0\leq G =\left(1-p_{e}\right)\bra{G}\rho_{g}\ket{G}\leq 1-p_{e}$}. Coherence in the energy ground space can be chosen to increase or decrease the thermalisation temperature $\beta$ of the system
\begin{equation}
\beta = \beta_{R} + \frac{1}{\hbar\omega}\ln\left(\ell \bra{G}\rho_{g}\ket{G}\right).
\end{equation}
In the case of no coherence then $\bra{G}\rho_{g}\ket{G} = \ell^{-1}$ and $\beta=\beta_{R}$. Coherence can be used to make the effective temperature arbitrarily hot; as $\bra{G}\rho_{g}\ket{G}$ approaches $\left(e^{-\beta_{R}\hbar\omega}\right)/\ell$ from above, the thermalisation temperature tends to infinity. However since $\bra{G}\rho_{g}\ket{G}\leq 1$, the effective temperature can only be lowered as cold as
\begin{equation}
\beta = \beta_{R} + \frac{1}{\hbar\omega}\ln(\ell).\label{eq:coollimit}
\end{equation}
A corollary is that when using maximal coherence to lower the effective temperature of a thermal bath, the final temperature is bounded from above
\begin{equation}
T \leq \frac{\hbar\omega}{\ln\left(\ell\right)}.
\end{equation}

It may seem paradoxical that the coherence in the ground state can raise the thermalisation temperature of the system arbitrarily; however, this is not a mysterious consequence of quantum coherence but simply due to the nature of the coupling between the atoms and field \eqref{eq:Hi}. By tuning the quantum state in the ground subspace $\rho_{g}$ we can prevent the system from releasing energy to the atoms and so each interaction can only increase the energy of the system.

For the rest of this paper we will specialise to the case in which $\ell=2$.

\section{The Otto cycle}\label{sec:Otto} 

\begin{figure}[t]
    \centering
    \includegraphics[width=0.45\textwidth]{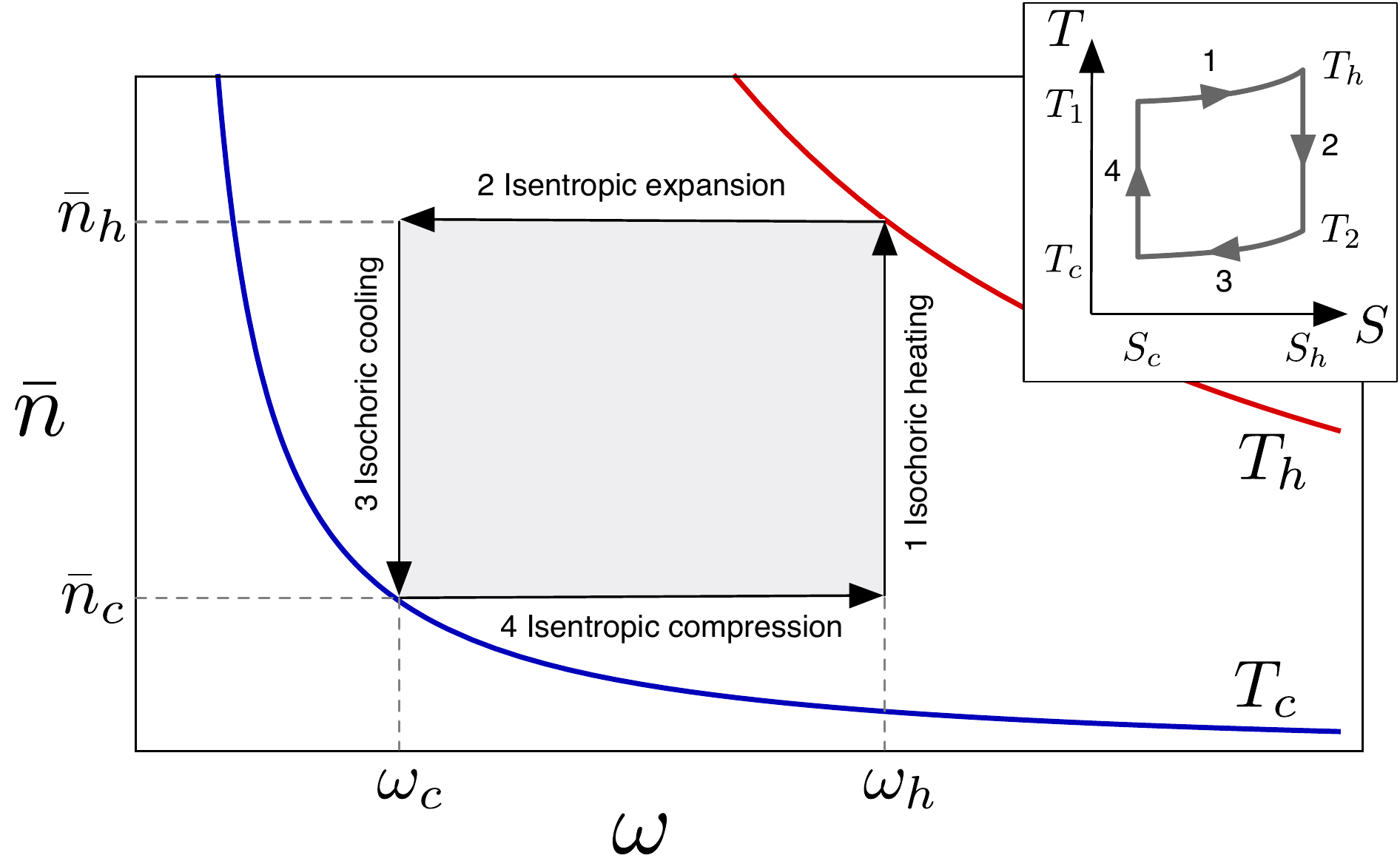}
    \caption{The Otto cycle shown for the variables $\bar{n}$ and $\omega$, and as $T$-$S$ diagram (inset). Strokes described in SEC.~\ref{sec:Otto}. Isotherms at temperatures $T_{h}$ (red) and $T_{c}$ (blue) are shown. The work extracted in a single cycle is the area enclosed by the rectangle, $W = \hbar\left(\omega_{h}-\omega_{c}\right)\left(\bar{n}_{h}-\bar{n}_{c}\right)$. The heat input to the system is the energy change of the system during the isochoric heating stroke, $Q_{h} = E_{1} = \hbar\omega_{h}\left(\bar{n}_{h}-\bar{n}_{c}\right)$.}
    \label{fig:ottocycle}
\end{figure}
In this section we examine the heat engine under the quasistatic evolution of the Otto  thermal cycle.
The Otto cycle (see FIG.~\ref{fig:ottocycle}) involves the following four strokes:
\begin{enumerate}
    \item  \emph{Isochoric heating}: the system interacts with the hot bath, eventually thermalising to temperature $T_h=\beta_{h}^{-1}$.
    \item  \emph{Isentropic expansion}: the system is isolated from the heat baths and the system frequency is decreased $\omega_{h}\rightarrow\omega_{c}$. Work is extracted from the system.
    \item  \emph{Isochoric cooling}: the system interacts with the cold bath, eventually thermalising to temperature $T_c=\beta_{c}^{-1}$.
    \item  \emph{Isentropic compression}: the system is isolated from the heat baths and the system frequency is increased $\omega_{c}\rightarrow\omega_{h}$. Work is performed on the system to close the cycle.
\end{enumerate}
The Otto cycle allows the easy quantification of heat and work for a cycle. Since the isochoric (constant volume) stokes involve no change of the system Hamiltonian, any change in energy is due to a heat flow between the system and thermal baths, $\Delta Q = \Delta E_{1}+\Delta E_{3}$. During the isentropic (constant entropy) strokes, the system is isolated, and so any change in energy is due to work done on the system, $-\Delta W = \Delta E_{2}+\Delta E_{4}$. Compare this to the Carnot cycle, where work is done on the system during every stroke. The internal energy is $U = \bar{H}_{S} = \hbar\omega \left( \bar{n} + \frac{1}{2} \right)$ and hence
\begin{equation}
    dU = \underbrace{\frac{dU}{d\omega}d\omega}_{-dW}+\underbrace{\frac{dU}{d\bar{n}}d\bar{n}}_{dQ},
\end{equation}
which is a statement of the first law for this system. During a quasi-static evolution the system remains in a thermal state. The entropy can be found by integrating
\begin{figure}[t]
    \centering
    \includegraphics[scale=0.4]{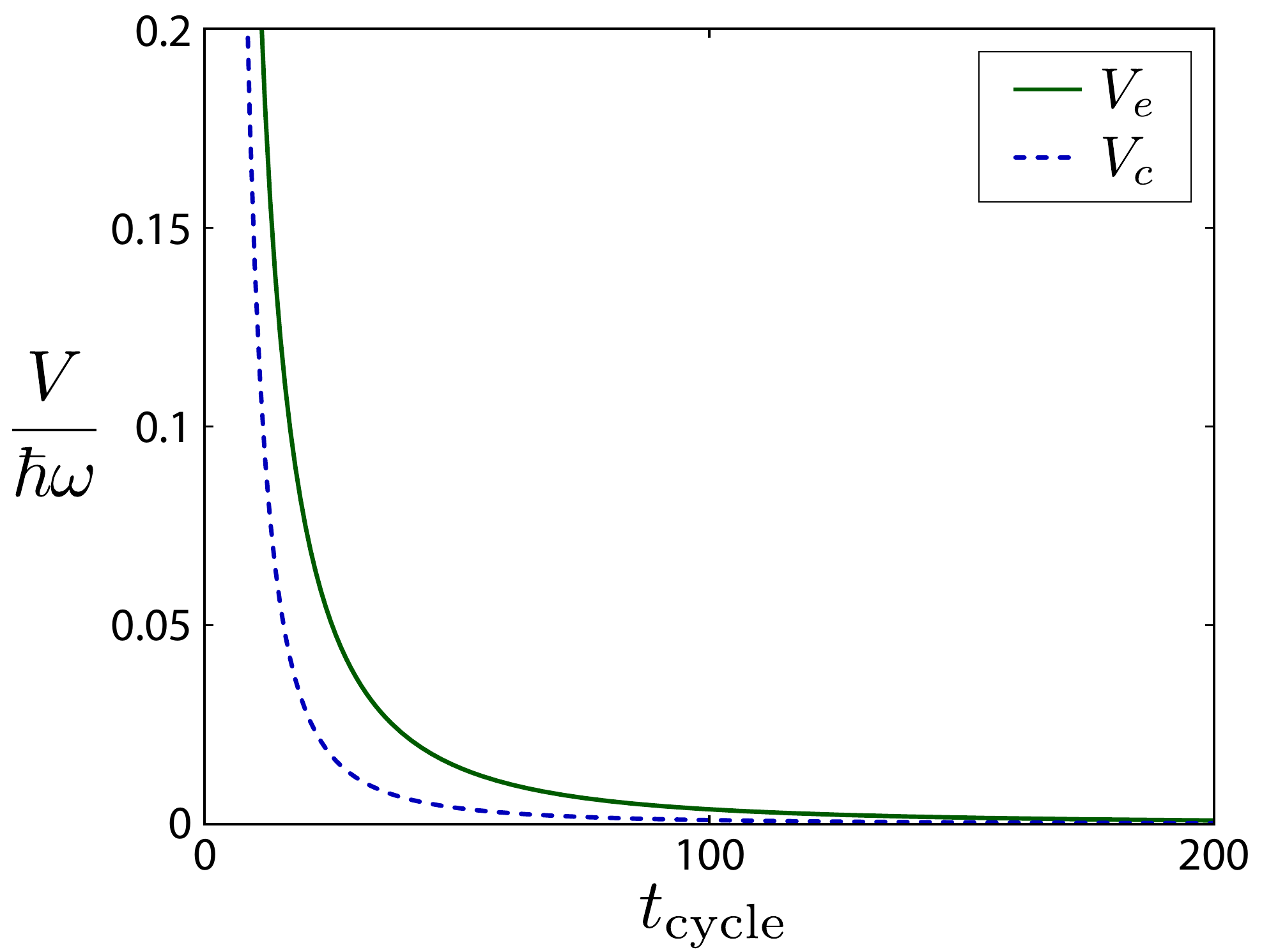}
    \caption{The energy cost of the isentropic expansion $V_{e}$ and compression $V_{c}$ strokes, using the shortcut-to-adiabaticity protocol. The energy cost is scaled by $1/\hbar\omega$, where $\omega = 10^{14}$Hz. Recall that $t_{\text{cycle}}$ is in terms of the dimensionless time parameter \eqref{eq:me}. The cost is large for short cycle times and approaches zero as they become longer in duration. In this plot we used a hot bath with temperature $\bar{n}_{\mathrm{ss}}=2$, and a cold bath with $\bar{n}_{\mathrm{ss}}=0.55$, both of which were incoherent. The system interacted with the hot bath for duration $t_{h}$ and with the cold bath for $t_{c}$; and we chose $t_{h}=t_{c}=t_{\text{cycle}}/4$. Expansion carries a larger cost than compression because it begins with the system at a hotter temperature, and $V\left(\tau\right)\propto \bar{n}\left(0\right)$ \eqref{eq:cost}. The compression stroke begins with the steady cycle initial photon number $\bar{n}_{sc}$ \eqref{eq:steadycycle}; the expansion stroke begins with initial photon number $\bar{n}_{h}$, which is found by evolving $\bar{n}_{sc}$ under \eqref{eq:nt} for time $t_{h}$. The integration in \eqref{eq:cost} was completed numerically.}
    \label{fig:expand_compress}
\end{figure}
\begin{equation}
    d S = \beta dQ = \ln\left(\frac{\bar{n}+1}{\bar{n}}\right) d\bar{n}
\end{equation}
from which we find
\begin{equation}
    S = \left(\bar{n}+1\right)\ln\left(\bar{n}+1\right)-\bar{n}\ln\left(\bar{n}\right).
\end{equation}
The efficiency of the Otto cycle is the ratio of the work extracted to the heat input from the hot bath
\begin{equation}
    \eta  \equiv\frac{W}{Q_{h}}=\frac{Q_{h}+Q_{c}}{Q_{h}}.
\end{equation}
We can easily read off from FIG.~\ref{fig:ottocycle} that the work done in a cycle is the area $W=\hbar\left(\omega_h-\omega_c\right)\left(\bar{n}_{h}-\bar{n}_{c}\right)$, and therefore 
\begin{equation}
    \eta = \frac{\hbar\left(\omega_{h}-\omega_{c}\right)\left(\bar{n}_{h}-\bar{n}_{c}\right)}{\hbar\omega_{h}\left(\bar{n}_{h}-\bar{n}_{c}\right)}=1-\frac{\omega_{c}}{\omega_{h}}.
\end{equation}
We can relate this to the temperature (see FIG.~\ref{fig:ottocycle}), since during the work strokes, the product $\beta\omega$ is constant,
\begin{equation}
    \eta = 1-\frac{T_{2}}{T_{h}} = 1-\frac{T_{c}}{T_{1}}.
\end{equation}

\section{The Finite Time Regime}\label{sec:finite}

Previous works \cite{2003scully862,Quan2006,Dillenschneider2009,Turkpence2016} have considered a harmonic oscillator undergoing a Carnot or Otto cycle in the quasistatic limit. However the quasistatic limit is impractical as it produces zero output power; so we examine the harmonic oscillator undergoing a finite time Otto cycle.
In this section we discuss two issues in operating a heat engine for finite time. During the work strokes, we use shortcut-to-adiabaticity techniques to control the system energy populations; and during the heat strokes, we need to guarantee that the average energy returns to its original value at the end of each cycle.

\subsection{Work Strokes}

During the work strokes, the system evolves under the time-dependent Hamiltonian \cite{EPLLutz2017,CHEMDeff2010,PTPHusimi1953}
\begin{equation}
    H\left(t\right) = \frac{1}{2}\left(\omega\left(t\right)^{2}x^{2}+p^{2}\right).
\end{equation}
If the system frequency is changed very slowly, then by the adiabatic theorem, the state of the system is constant. Quickly changing the frequency will excite off-diagonal terms in the density matrix. Using \emph{shortcuts-to-adiabaticity} (STA) techniques \cite{ATOMTorrontegui2013}, the state of the system can be constant even when quickly changing the frequency (this has recently been demonstrated experimentally \cite{Deng2018,Diao2018}). This can be achieved by adding a \emph{counterdiabatic term} to the Hamiltonian, $H\left(t\right)\rightarrow H\left(t\right)+H_{\text{STA}}\left(t\right)$. For example, in \cite{DelCampo2013}, the author uses the following counterdiabatic term
\begin{equation}
    H_{\text{STA}}\left(t\right)=\frac{1}{2}\left(\frac{\ddot{\omega}}{2\omega}-\frac{3\dot{\omega}^{2}}{4\omega^{2}}\right)x^{2}.\label{eq:conterdiabatic}
\end{equation}
The author changes the system frequency according to
\begin{align}
    \omega\left(t\right) =& \;\omega_{i} +10\left(\omega_{f}-\omega_{i}\right)\left(\frac{t}{\tilde{\tau}}\right)^{3}-15\left(\omega_{f}-\omega_{i}\right)\left(\frac{t}{\tilde{\tau}}\right)^{4}\nonumber\\
    &+6\left(\omega_{f}-\omega_{i}\right)\left(\frac{t}{\tilde{\tau}}\right)^{5}.
\end{align}
This protocol will change the frequency from $\omega_{i}\rightarrow\omega_{f}$ in time $\tilde{\tau}$. The author shows that if the system begins in a thermal state, then under this counterdiabatic Hamiltonian \eqref{eq:conterdiabatic}, the energy populations will be constant. Such a protocol can be used to drive an thermal cycle \cite{Campo2015,Beau2016,Abah2018,Abah2018PRE}; we use this protocol as well.  The authors in \cite{EPLLutz2017} describe an energy cost associated with the addition of this counterdiabatic term, given by
\begin{align}
   V\left(\tilde{\tau}\right) &=  \frac{1}{\tilde{\tau}}\int_{0}^{\tilde{\tau}} \expect{H_{\text{STA}}\left(t\right)}dt\nonumber \\
   &=\frac{\hbar \bar{n}\left(0\right)}{\tilde{\tau}}\int_{0}^{\tilde{\tau}}\omega\left(\frac{\ddot{\omega}}{4\omega^{3}}-\frac{\dot{\omega}^{2}}{4\omega^{4}}\right)dt, \label{eq:cost}
\end{align}
where $\bar{n}\left(0\right)$ is the average photon number at $t=0$; which implies the isentropic expansion strokes will have a larger energy cost than the compression strokes (see FIG.~\ref{fig:expand_compress}; in all our plots, the integration was completed numerically). The energy cost is monotonically decreasing in the stroke duration. We note here that while we adopt the energy cost described in \cite{EPLLutz2017}, there exists a diversity of opinion about what constitutes the appropriate energy cost of running a shortcut-to-adiabaticity protocol \cite{Abah2018,Kosloff2017,Torrontegui2017,Zheng2016,Calzetta2018}.

The efficiency of the heat engine will therefore be diminished by two energy costs $V_{e}$ and $V_{c}$, one each for the isentropic expansion and compression strokes
\begin{equation}
    \eta = \frac{W-V_{e}-V_{c}}{Q_{h}}. \label{eq:efficiencywithcost}
\end{equation}

The authors of \cite{EPLLutz2017} consider the cost $V_{e}+V_{c}$ as an additional heat input into the system. We disagree with this. This energy cost is due to a unitary process when the system is isolated from other systems. Therefore we consider the energy cost as a reduction in the work extracted from the system.

Using this counterdiabatic term the energy populations of the system are constant, and therefore the average photon number $\bar{n}$ of the system is constant. We assume that the change in energy of the system during the isentropic expansion stroke $\Delta\bar{H}_{S}=\hbar\left(\omega_{h}-\omega_{c}\right)\bar{n}_{h}$ is entirely extracted as work. Hence, the net extracted work is proportional to the change in the frequency of the system $\omega_{h} - \omega_{c}$.

\subsection{Heat Strokes}

\begin{figure}[t]
    \centering
    \includegraphics[scale=0.38]{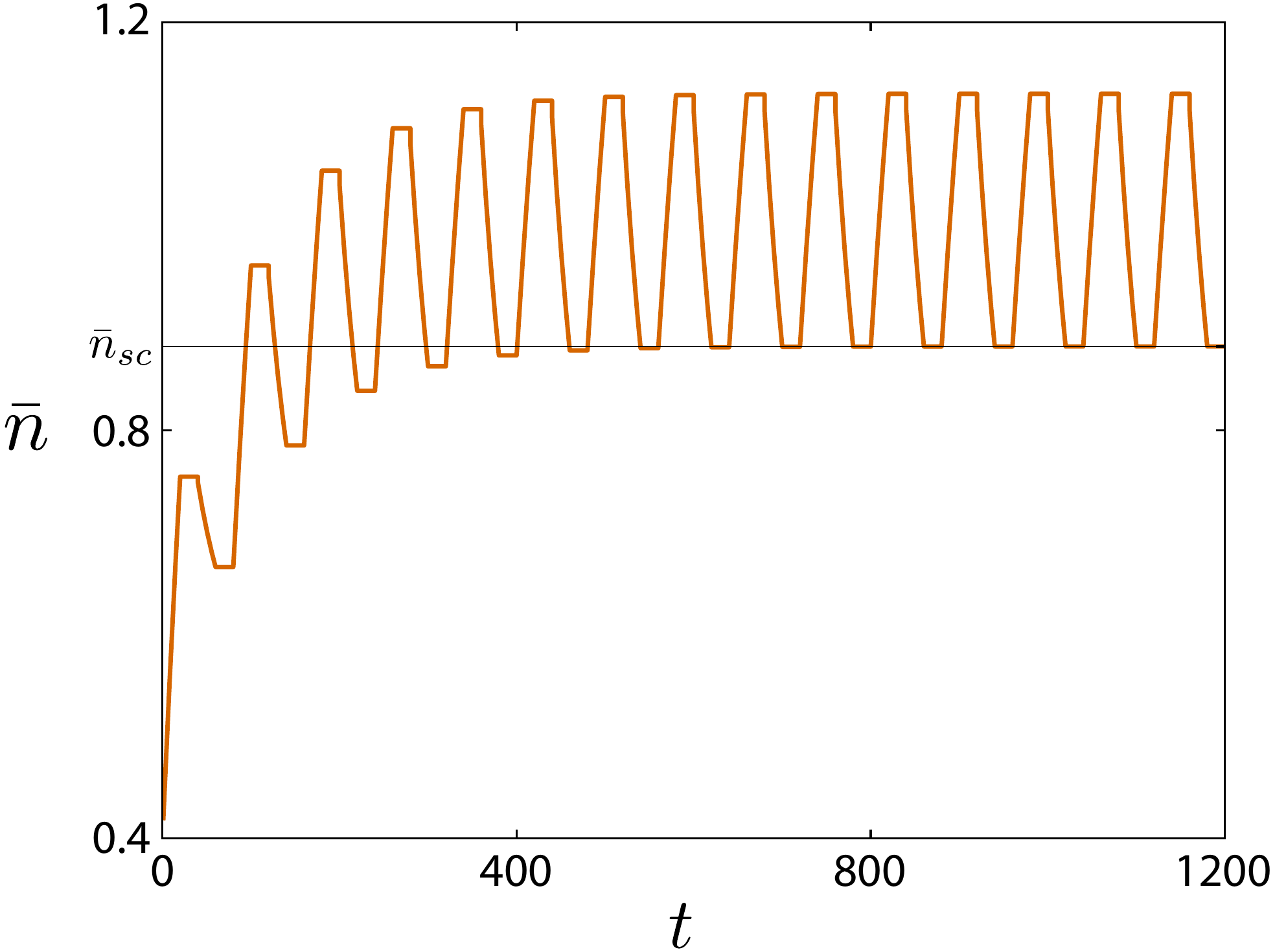}
    \caption{The average photon number $\bar{n}$ as the system reaches a steady cycle. Regardless the initial average photon number $\bar{n}_{0}$, $\bar{n}$ eventually reaches a closed cycle \eqref{eq:steadylimit}. The photon number is constant during the work strokes, since the shortcut-to-adiabaticity protocol holds the energy populations constant. Each stroke takes a quarter of the total cycle time: $t_{\text{cycle}}/4$.}
    \label{fig:steady_cycle}
\end{figure}

During the heating (isochoric) strokes, the system frequency $\omega$ is constant, and the heat into the system is proportional to the change in the average photon number  $\bar{n}_{h}-\bar{n}_{c}$ of the system.

We assume the system begins in a thermal state. After a single cycle, the system has interacted with the hot bath for time $t_{h}$, and the cold bath for time $t_{c}$. If the systems begins with an arbitrary initial average photon number $\bar{n}_{0}$, it is not guaranteed that the cycle will be closed, since $\bar{n}$ may not return to its initial value. In this section we show that the system will always asymptotically approach a ``steady cycle", where the average photon number at the end of the cycle is the same as that at the beginning of the cycle.

Since $\bar{n}$ is constant during the work strokes, the average photon number after one cycle is given by two applications of \eqref{eq:nt},
\begin{align}
    \bar{n}_{1} = &\left(\left(\bar{n}_{0}-\frac{E_{h}}{\Delta_{h}}\right)e^{-\Delta_{h}t_{h}}+\frac{E_{h}}{\Delta_{h}}-\frac{E_{c}}{\Delta_{c}}\right) \nonumber\\
    &\times e^{-\Delta_{c}t_{c}} + \frac{E_{c}}{\Delta_{c}}.
    \label{eq:aftercycle}
\end{align}
Recall that we defined $\Delta \equiv G-E$, the difference in the transition rates of the master equation \eqref{eq:me}.
We have introduced subscripts $h$ and $c$ to denote the terms which arise due to interactions with the hot and cold baths respectively.
We want to know the initial photon number such that $\bar{n}_{1}=\bar{n}_{0}$. We denote it $\bar{n}_{sc}$,
\begin{align}
\bar{n}_{sc} &= \frac{\frac{E_{h}}{\Delta_{h}}e^{-\Delta_{c}t_{c}}\left(e^{-\Delta_{h}t_{h}}-1\right)+\frac{E_{c}}{\Delta_{c}}\left(e^{-\Delta_{c}t_{c}}-1\right)}{e^{-\Delta_{h}t_{h}}e^{-\Delta_{c}t_{c}}-1}. \nonumber \\
&= \frac{\frac{E_{h}}{\Delta_{h}}e^{\frac{-\Delta_{c}t_{c}}{2}}\sinh\left(\frac{\Delta_{h}t_{h}}{2}\right)+\frac{E_{c}}{\Delta_{c}}e^{\frac{\Delta_{h}t_{h}}{2}}\sinh\left(\frac{\Delta_{c}t_{c}}{2}\right)}{\sinh\left(\frac{\Delta_{h}t_{h}+\Delta_{c}t_{c}}{2}\right)}.
\label{eq:steadycycle}
\end{align}

If the interaction time with the cold bath $t_{c}$ is large, then $\bar{n}_{sc}$ is the thermalisation average photon number of the cold bath.
\begin{equation}
\lim_{t_{c}\rightarrow\infty} \bar{n}_{sc} = \frac{E_{c}}{\Delta_{c}}.
\end{equation}
So if the system begins with $\bar{n}_{0}=n_{sc}$, then after one cycle, the average photon number will return to $\bar{n}_{0}$. However even if $\bar{n}_{0}\ne \bar{n}_{sc}$, after multiple cycles the the average photon number at the beginning of a cycle will tend to $\bar{n}_{sc}$. After $\delta$ cycles (see Appendix~\ref{sec:Steady_cycle} for details)
\begin{equation}
    \bar{n}_{\delta} = e^{-\delta\Delta_{h}t_{h}}e^{-\delta\Delta_{c}t_{c}}\left(\bar{n}_{0}-\bar{n}_{sc}\right)+\bar{n}_{sc},
\end{equation}
so, as seen in FIG.~\ref{fig:steady_cycle},
\begin{equation}
    \lim_{\delta\rightarrow\infty} \bar{n}_{\delta} = \bar{n}_{sc},\label{eq:steadylimit}
\end{equation}
and the system always reaches a `steady cycle', where the average photon number returns to its initial value after each cycle. This has been previously studied in \cite{SCFeldmann1996,Kosloff2017}. We assume for the rest of this paper that the system has reached a steady cycle.

\section{Performance}\label{sec:results}

\subsection{The effect of coherence}
The work extracted in a single cycle is $\hbar\left(\omega_{h}-\omega_{c}\right)\left(\bar{n}_{h}-\bar{n}_{c}\right)$. We can use \eqref{eq:nt} to write
\begin{equation}
\bar{n}_{h}-\bar{n}_{c} = \left(\bar{n}_{c}-\frac{E_{h}}{\Delta_{h}}\right)\left(e^{-\Delta_{h} t_{h}}-1\right),
\end{equation}
where the average photon number begins from $\bar{n}\left(0\right)=\bar{n}_{c}$ and evolves to $\bar{n}\left(t_{h}\right)=\bar{n}_{h}$.
In a steady cycle, $\bar{n}_{c}$ is given by \eqref{eq:steadycycle}, and so
\begin{align}
\bar{n}_{h}-\bar{n}_{c} = \left(\frac{E_{h}}{\Delta_{h}}-\frac{E_{c}}{\Delta_{c}}\right)\frac{2\sinh\left(\frac{\Delta_{c}t_{c}}{2}\right)\sinh\left(\frac{\Delta_{h}t_{h}}{2}\right)}{\sinh\left(\frac{\Delta_{c}t_{c}+\Delta_{h}t_{h}}{2}\right)}. \label{eq:nh-nc}
\end{align}
Keeping the temperature of the baths constant, \mbox{$\bar{n}_{h}-\bar{n}_{c}$} is monotonically increasing in both $\Delta_{h}$ and $\Delta_{c}$ (see Appendix~\ref{sec:AppWork}, for the calculation).

We have seen that coherence in the energy ground space of the incoming atoms can increase or decrease the effective temperature of atomic thermal bath (see section~\ref{sec:interaction}).
To investigate the effect of coherence on the power of the engine, we study three pairs (hot and cold) of heat baths, all at the same effective inverse temperatures, $\beta_{h}$ and $\beta_{c}$. The first is a pair of incoherent (\textsc{I}) baths, with thermalisation photon numbers
\begin{equation}
\begin{matrix}
\dfrac{E_{h}^{\textsc{i}}}{\Delta_{h}^{\textsc{i}}} &= \dfrac{1}{e^{\beta_{h}\hbar\omega}-1}, \\[12pt]
\dfrac{E_{c}^{\textsc{i}}}{\Delta_{c}^{\textsc{i}}} &= \dfrac{1}{e^{\beta_{c}\hbar\omega}-1}.
\end{matrix}
\end{equation}
\begin{figure}[b]
\includegraphics[scale=0.38]{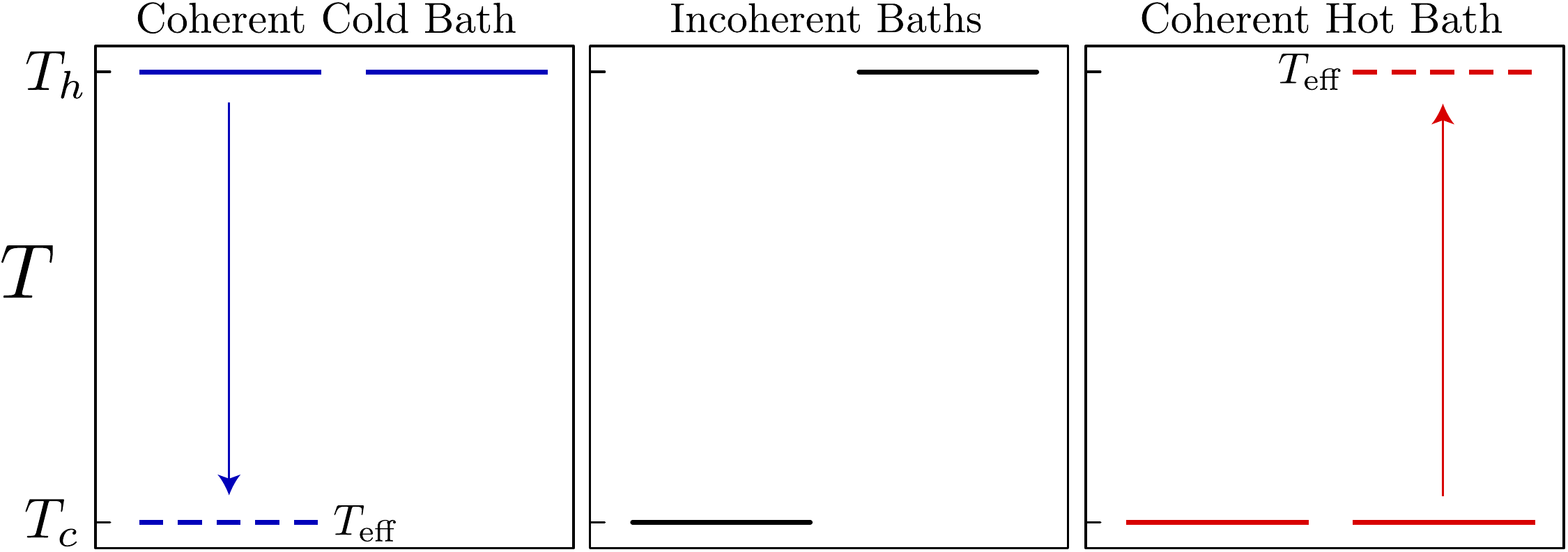}
\begin{tabular}{l | c | c | c | c }
Incoherent Baths (\textsc{I}) & $E_{h}$ & $\Delta_{h}$ & $E_{c}$ & $\Delta_{c}$ \\ \hline
Coherent Hot Bath (\textsc{CH}) & $E_{c}$ & $\Delta_{h}\frac{E_{h}}{E_{c}}$ & $E_{c}$ & $\Delta_{c}$ \\ \hline
Coherent Cold Bath (\textsc{CC}) & $E_{h}$ & $\Delta_{h}$ & $E_{h}$ & $\Delta_{c}\frac{E_{c}}{E_{h}}$
\end{tabular}
\caption{Diagram and table of three different pairs of baths, all with the same effective temperatures. Temperature according to thermal populations in hard line, effective temperature in broken lines. The coherent hot (\textsc{CH}) baths involve using coherence to create a hot bath from two cold baths; the coherent cold (\textsc{CC}) baths use coherence to create a cold bath from two hot baths. Without the use of coherence, these latter two bath systems would not be usable for running a thermal engine.}
\label{fig:bath_table}
\end{figure}
We compare this with two other pairs of baths with the same effective temperatures. The first pair (coherent hot or \textsc{CH}) have cold thermal populations consistent with inverse temperature $\beta_{c}$, however there is coherence in the energy ground space of one of the baths such that the effective inverse temperature is that of the incoherent hot bath $\beta_{h}$. Explicitly
\begin{equation}
E_{h}^{\textsc{ch}}=E_{c}^{\textsc{ch}}=E_{c}^{\textsc{i}},\quad \Delta_{c}^{\textsc{ch}} = \Delta_{c}^{\textsc{i}},\quad \Delta_{h}^{\textsc{ch}} = \Delta_{h}^{\textsc{i}}\frac{E_{c}^{\textsc{i}}}{E_{h}^{\textsc{i}}}.
\end{equation}
The second pair of thermal baths (coherent cold or \textsc{CC}) is the reverse: hot  thermal populations at inverse temperature $\beta_{h}$, but with sufficient coherence in the ground space of one of the bath such that it has an effective inverse temperature of the incoherent cold bath $\beta_{c}$. That is 
 \begin{equation}
E_{h}^{\textsc{cc}}=E_{c}^{\textsc{cc}}=E_{h}^{\textsc{i}},\quad \Delta_{h}^{\textsc{cc}} = \Delta_{h}^{\textsc{i}},\quad \Delta_{c}^{\textsc{cc}} = \Delta_{c}^{\textsc{i}}\frac{E_{h}^{\textsc{i}}}{E_{c}^{\textsc{i}}}.
\end{equation}
These three different pairs of baths are summarised in FIG.~\ref{fig:bath_table}.
Without the addition of coherence, these latter two pairs of baths would be unusable; the addition of coherence creates an effective temperature difference, and allows them to be used to extract work, as seen in \cite{2003scully862}. In the first case the coherence is being used to turn a cold bath into a hot bath, and in the second case coherence is being used to change a hot bath into a cold bath. We note here that
\begin{equation}
\Delta_{h}^{\textsc{ch}} < \Delta_{h}^{\textsc{i}}\text{ and } \Delta_{c}^{\textsc{cc}} > \Delta_{c}^{\textsc{i}}.
\label{eq:newbaths}
\end{equation}
From \eqref{eq:nt}, $\Delta=G-E$ effects not only the effective temperature of the bath but also the rate at which the system thermalises. Faster thermalisation allows more work to be extracted per cycle. Creating the coherent hot bath requires a smaller $\Delta_{h}$ as compared with the incoherent bath pair, and less work will be extracted in a cycle. Conversely, the creation of the coherent cold bath requires a larger $\Delta_{c}$ than the incoherent baths, and more work will be extracted per cycle.

\subsection{Cost}\label{sec:cost}

\begin{figure}[t]
\includegraphics[scale=0.4]{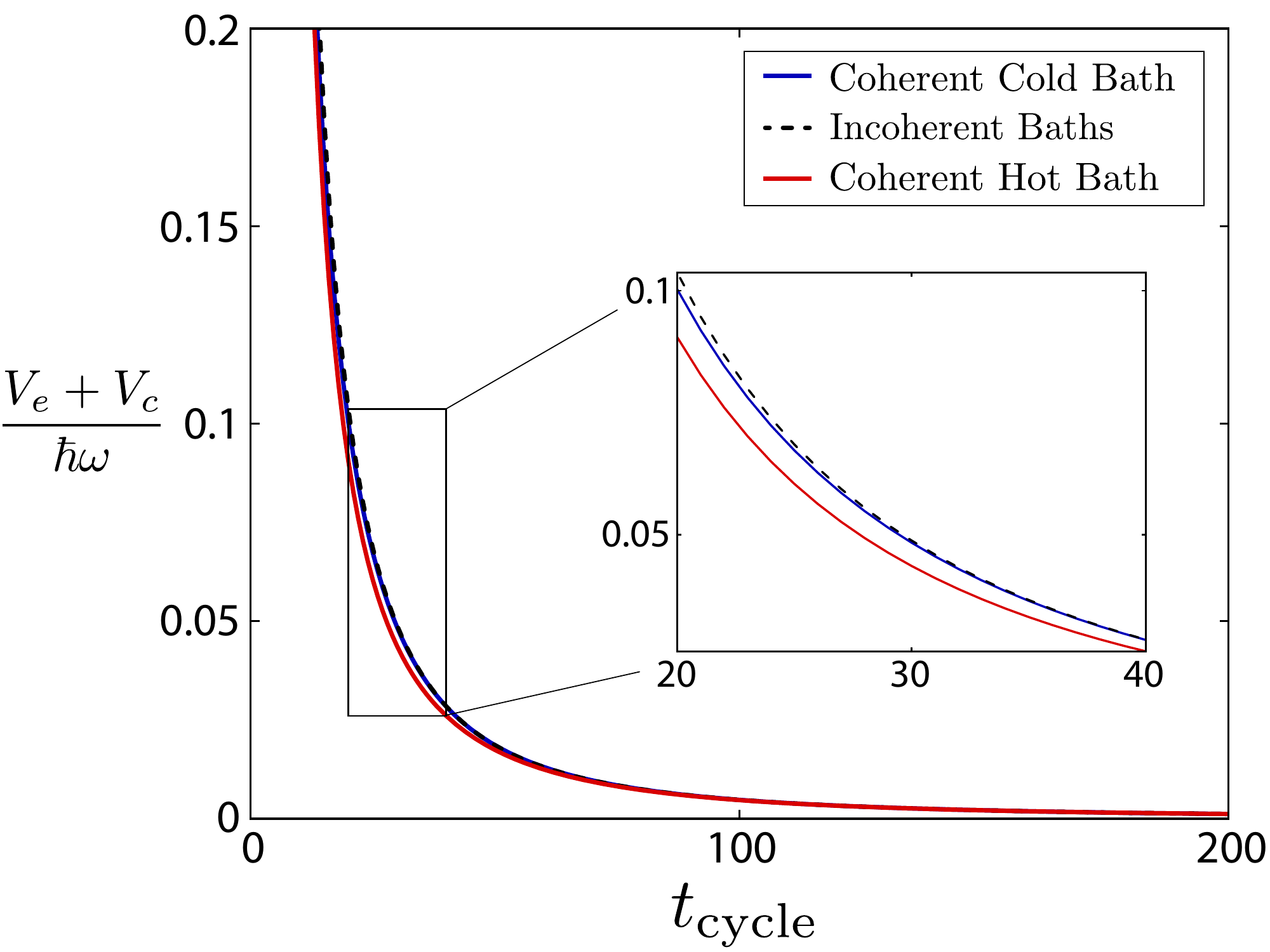}
\caption{The cost of the shortcut to adiabaticity protocol $V_{e}+V_{c}$ (see \eqref{eq:cost} and FIG.~\ref{fig:expand_compress}) for the three pairs of thermal baths. The total energy cost is scaled by $1/\hbar\omega$, where $\omega= 10^{14}$Hz. Recall that $t_{\text{cycle}}$ is in terms of the dimensionless time parameter \eqref{eq:me}. The three pairs of thermal baths are described in FIG.~\ref{fig:bath_table}. Both the coherent cold \textsc{CC} and the coherent hot \textsc{CH} systems have a reduced cost derived from their coherence (see Appendix~\ref{sec:AppCost} for details). The cost is large for $t_{\text{cycle}}\rightarrow 0$. As $t_{\text{cycle}}\rightarrow\infty$, the cost $V_{e}+V_{c}\rightarrow 0$. Here we used a hot bath with temperature $\bar{n}_{\mathrm{ss}}=2$, and a cold bath with $\bar{n}_{\mathrm{ss}}=0.55$.}\label{fig:cost}
\end{figure}

The cost of the shortcut-to-adiabaticity protocol \eqref{eq:cost} is proportional to the average photon number $\bar{n}$ of the system at the beginning of the protocol. This will be different for the different bath systems, since this depends not only on the temperature of the baths but also on the thermalisation rates $\Delta_{h}$ and $\Delta_{c}$. The cost of the compression stroke $V_{c}$ is proportional to the average photon number $\bar{n}_{c}$. Since we are assuming the system has reached a steady cycle, this is given by $\bar{n}_{sc}$ \eqref{eq:steadycycle}. If we keep the temperature of the baths constant but change the absorption rate $\Delta_{h}$ and $\Delta_{c}$, we can see that $\bar{n}_{sc}$ is monotonically increasing in $\Delta_{h}$, but monotonically decreasing in $\Delta_{c}$. This is also true for the expansion stroke cost $V_{e}$, during which the average photon number is $\bar{n}_{h}$ (see Appendix~\ref{sec:AppCost} for details).\\
This implies, from \eqref{eq:newbaths} that the use of coherence either to increase the temperature of a hot bath, or to decrease the temperature of a cold bath, both reduce the cost of the shortcut-to-adiabaticity protocol. As can be seen in FIG.~\ref{fig:cost}, the \textsc{CC} and \textsc{CH} bath systems using coherence have a smaller energy cost than the incoherent (\textsc{I}) bath system.

\subsection{Efficiency}\label{sec:efficiency}

\begin{figure}[t]
    \centering
    \includegraphics[scale=0.4]{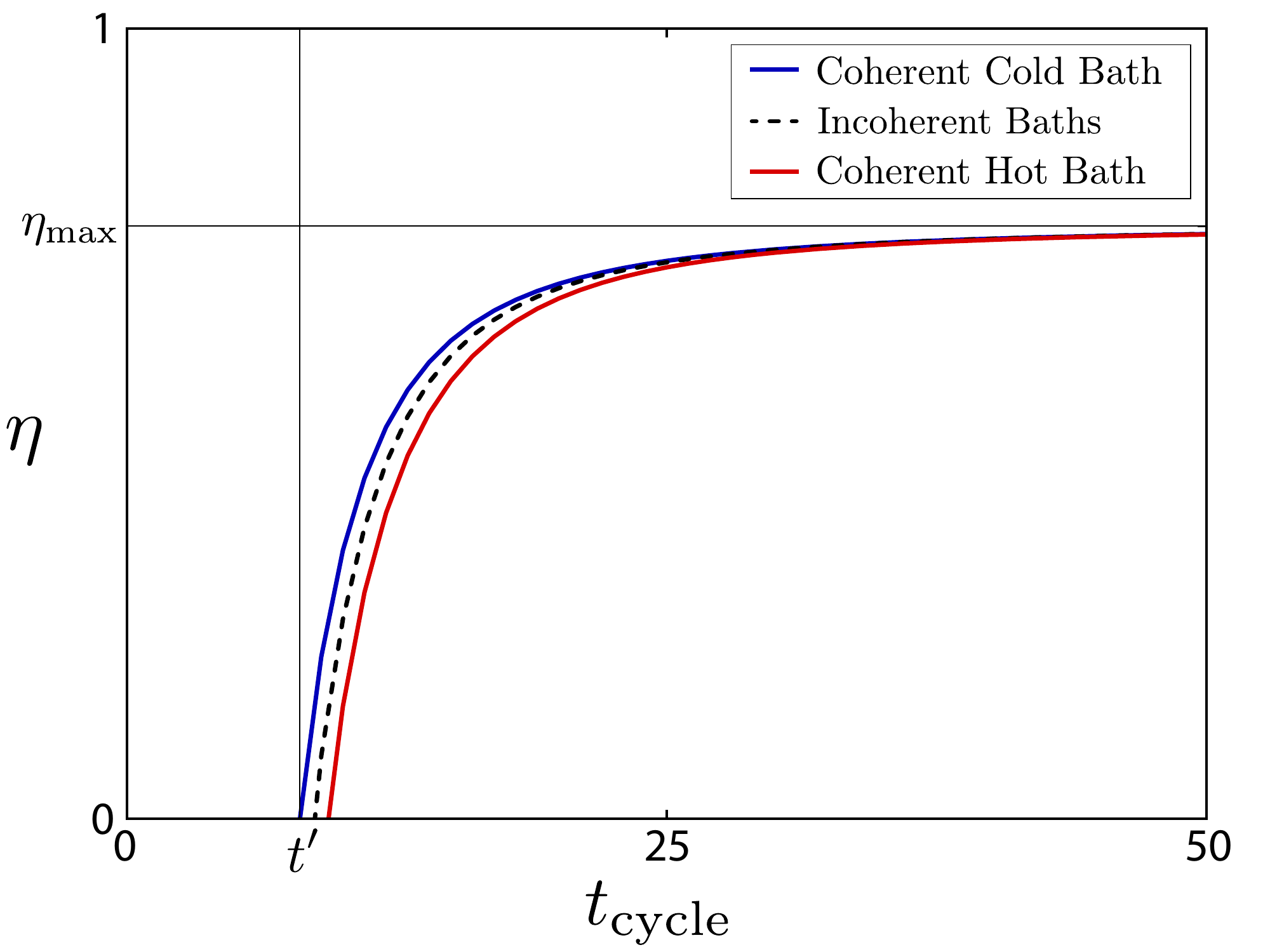}
    \caption{Efficiency curves \eqref{eq:efficiency} of the different baths. Recall that $t_{\text{cycle}}$ is in terms of the dimensionless time parameter \eqref{eq:me}. The three pairs of thermal baths are described in FIG.~\ref{fig:bath_table}. The coherent cold \textsc{CC} always has the highest efficiency \eqref{eq:effCC}. The coherent hot \textsc{CH} has the worst \eqref{eq:effCH}. Before $t^{\prime}$, the work extracted is less than the energy cost of the shortcut-to-adiabaticity protocol $W<V_{e}+V_{c}$. With longer cycle times, the cost of implementing the shortcut-to-adiabaticity protocol becomes negligible, and the efficiency approaches its maximum. Here we used a hot bath with temperature $\bar{n}_{\mathrm{ss}}=2$, and a cold bath with $\bar{n}_{\mathrm{ss}}=0.55$.}
    \label{fig:efficiency}
\end{figure}

The efficiency \eqref{eq:efficiencywithcost} is reduced from maximum because of the cost of implementing the shortcut-to-adiabaticity protocol.
\begin{equation}
\eta = \frac{W-V_{e}-V_{c}}{Q_{h}} = 1-\frac{\omega_{c}}{\omega_{h}}-\frac{V_{e}+V_{c}}{\hbar\omega_{h}\left(\bar{n}_{h}-\bar{n}_{c}\right)}.\label{eq:efficiency}
\end{equation}
Since the cost tends toward zero with increasing cycle time $t_{\text{cycle}}$, engines in which the expansion and compression strokes have a longer duration will be more efficient (as shown in FIG.~\ref{fig:efficiency}). There is a time $t^{\prime}$, before which, the energy cost of the shortcut-to-adiabaticity protocol will exceed the work extracted. This will result in a negative efficiency, rendering the engine too expensive to run.

\begin{figure}[t]
\includegraphics[scale=0.4]{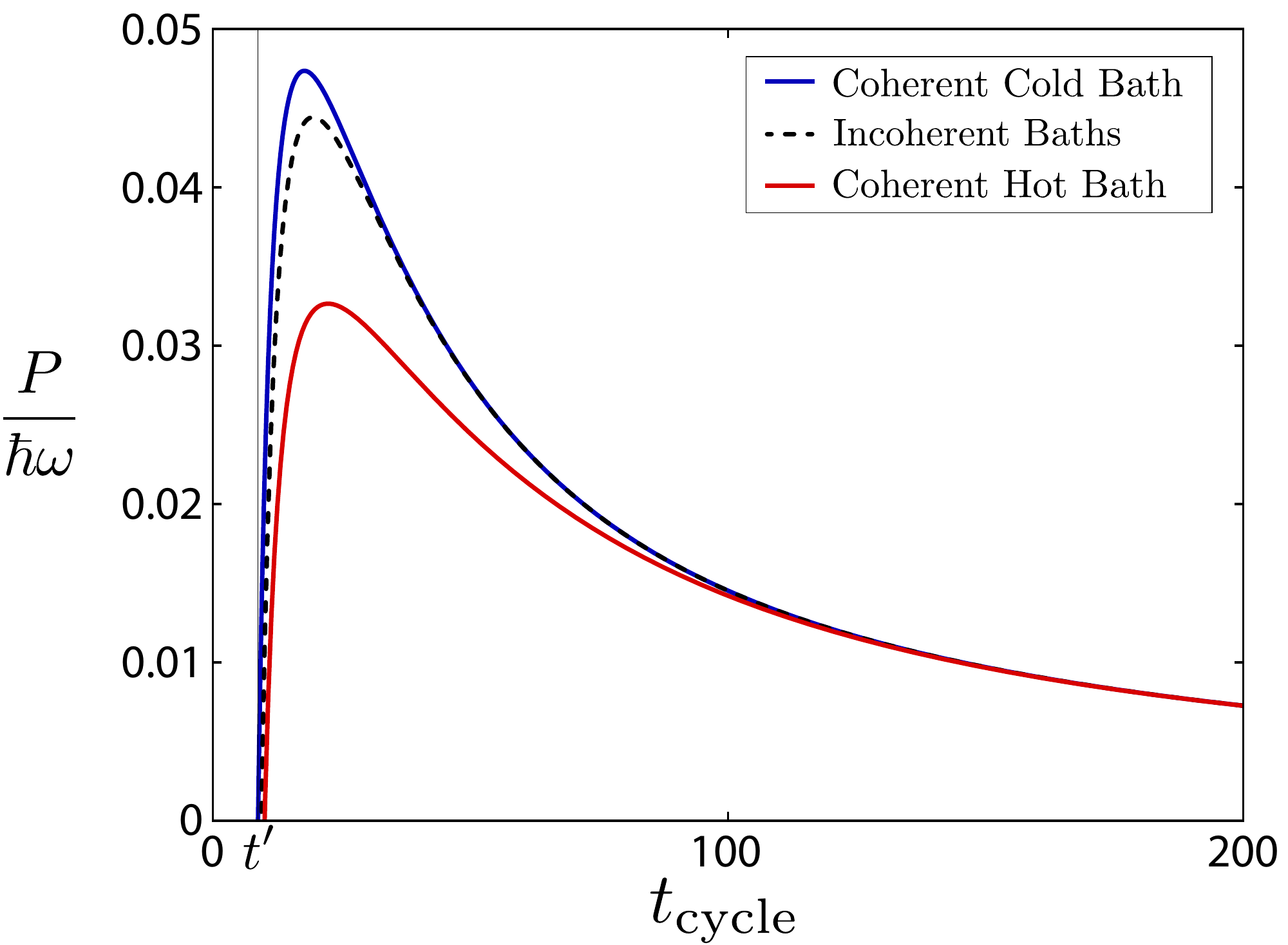}
\caption{The power curves \eqref{eq:power} for the three pairs of heat baths. The power is scaled by $1/\hbar\omega$, where $\omega= 10^{14}$Hz. Recall that $t_{\text{cycle}}$ is in terms of the dimensionless time parameter \eqref{eq:me}. The three pairs of thermal baths are described in FIG.~\ref{fig:bath_table}. The \textsc{CC} baths always have a higher output power than the incoherent baths \eqref{eq:powerCC}. The \textsc{CH} baths do not necessarily have a lesser output, but they do for these temperatures (see Appendix~\ref{sec:AppPower} for details). As $t_{h},t_{c}\rightarrow\infty$, the difference $\bar{n}_{h}-\bar{n}_{c}\rightarrow\frac{E_{h}}{\Delta_{h}}-\frac{E_{c}}{\Delta_{c}}$, $W$ approaches a constant value and $P\rightarrow 0$. Before $t^{\prime}$, the engine is too costly to run as the cost of using shortcut-to-adiabaticity techniques is larger than the work extracted $W<V_{e}+V_{c}$. Here we used a hot bath with temperature $\bar{n}_{\mathrm{ss}}=2$, and a cold bath with $\bar{n}_{\mathrm{ss}}=0.55$.}
\label{fig:power}
\end{figure} 

For the baths that we have hitherto examined, coherence has reduced the cost of the shortcut-to-adiabaticity protocol. However the reduction in the efficiency is also inversely dependent upon the heat input to the system by the hot bath $Q_{h}$, which is proportional to $\bar{n}_{h}-\bar{n}_{c}$ \eqref{eq:nh-nc}. We saw previously that $\bar{n}_{h}-\bar{n}_{c}$ is monotonically increasing in both $\Delta_{h}$ and $\Delta_{c}$ (see Appendix~\ref{sec:AppWork}). In the case of the $\textsc{CC}$ thermal baths the thermalisation rate $\Delta^{\textsc{cc}}_{c}>\Delta^{\textsc{i}}_{c}$, so the heat input from the hot bath $Q_{h}^{\textsc{cc}}$, is greater than the heat input using incoherent thermal baths $Q_{h}^{\textsc{i}}$,
\begin{equation}
Q_{h}^{\textsc{cc}} > Q_{h}^{\textsc{i}}.
\end{equation}
\begin{figure*}[t]
\subfigure[]{\includegraphics[scale=0.29]{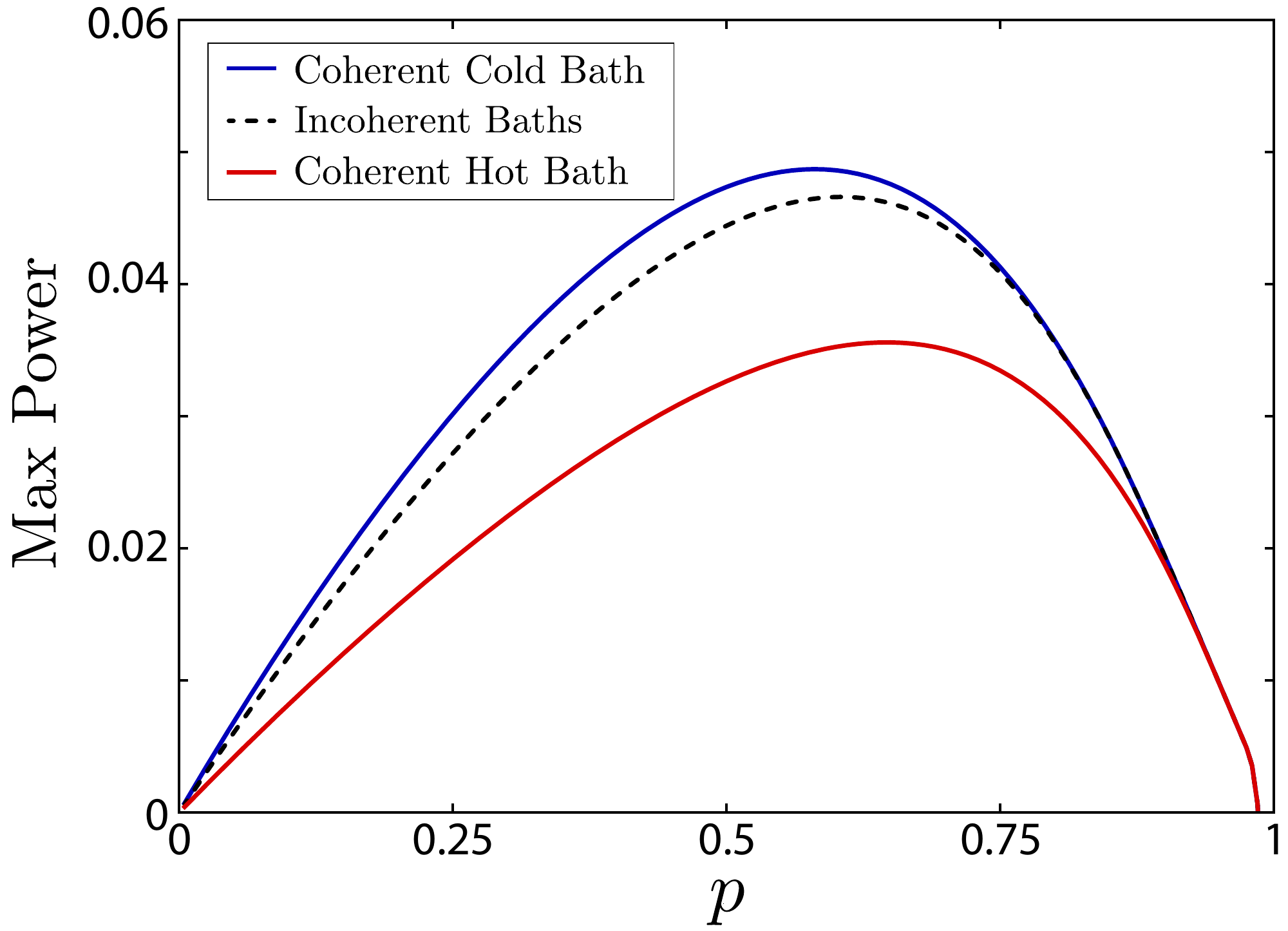}\label{fig:maxp}}
\subfigure[]{\includegraphics[scale=0.29]{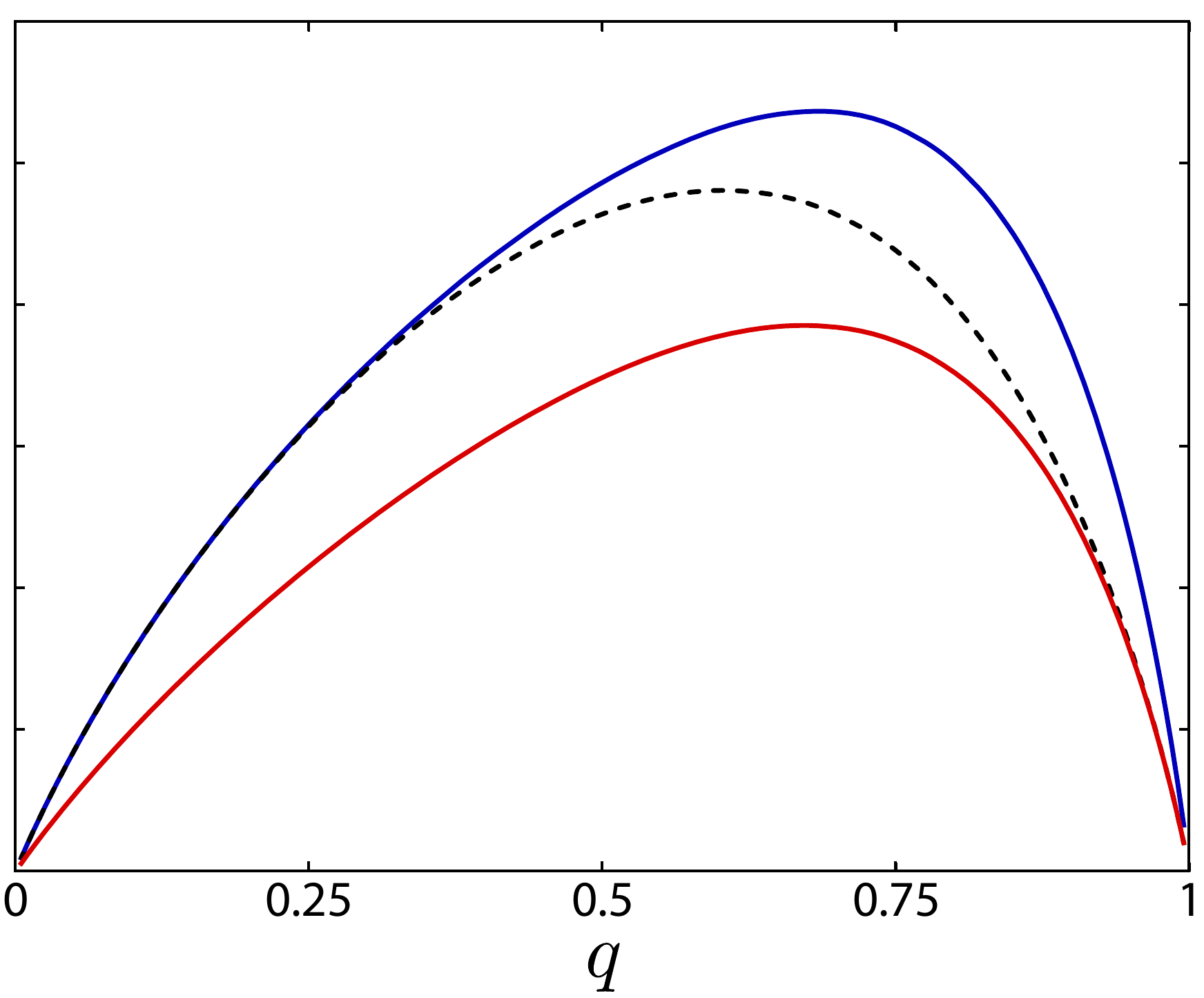}}
\subfigure[]{\includegraphics[scale=0.29]{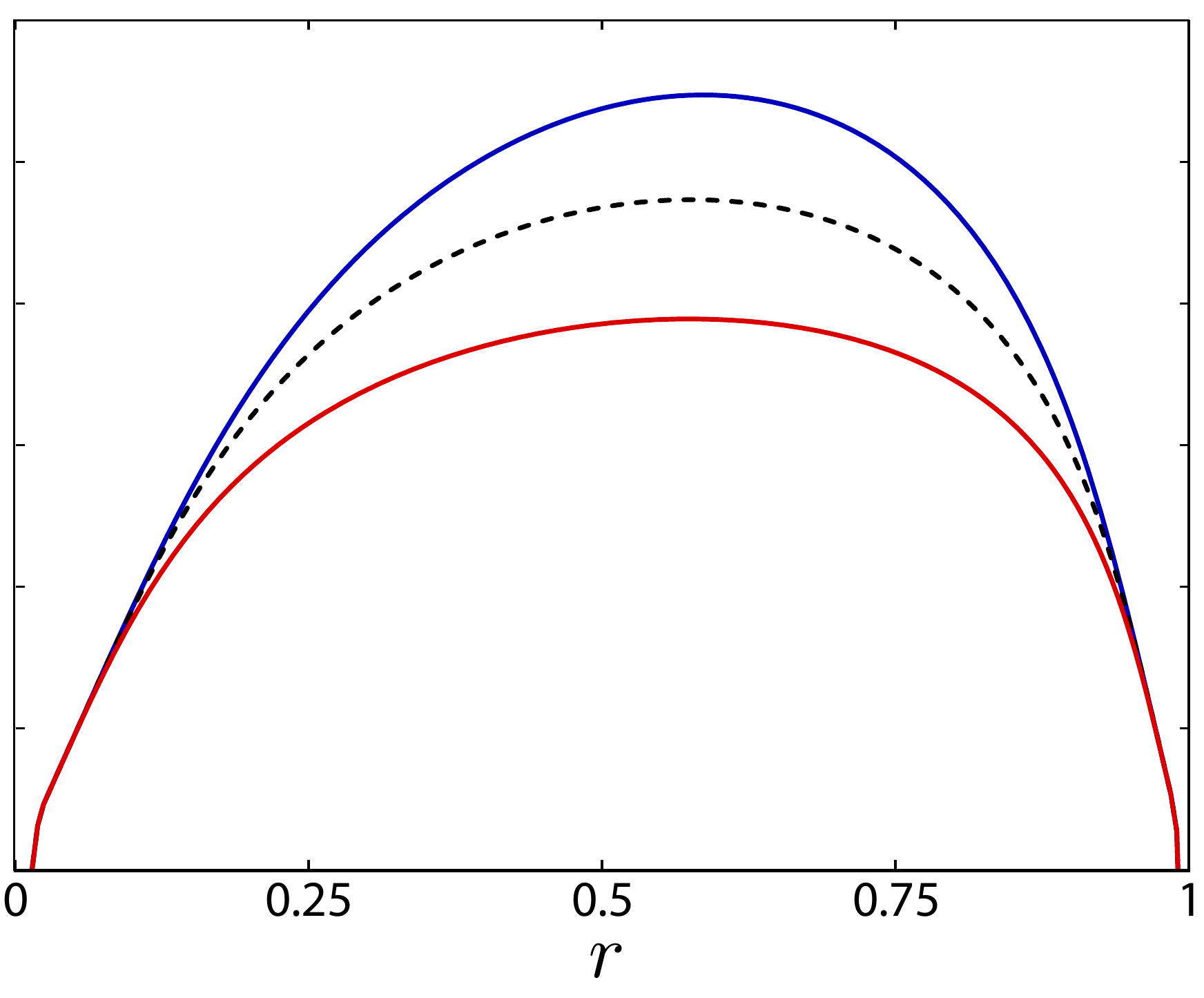}}
\caption{The power curves of the engines can be increased by changing the relative times of the four strokes of the Otto cycle. This will change the amount of heat transfer to the system in each cycle, and alter the energy cost of using shortcut-to-adiabaticity techniques. The power is scaled by $1/\hbar\omega$, where $\omega= 10^{14}$Hz. The three pairs of thermal baths are described in FIG.~\ref{fig:bath_table}. The cycle time $t_{\text{cycle}}$ is parametrised in \eqref{eq:Ptcycle} and \eqref{eq:paramqr}. Here we numerically find the optimal values of $p,q,r$ for the \text{CC} baths. These are three cross-sections of the function Max Power $(p,q,r)$, one for each variable \eqref{eq:maxpqr}. In (a), $q=r=0.5$, and the peak occurs at $p\approx 0.58$ for the \textsc{CC} baths. In (b), $p=0.58$, $r=0.5$ and the peak occurs at $q\approx 0.69$. In (c), $p=0.58, q=0.69$ and the peak occurs at $r\approx 0.59$. See FIG.~\ref{fig:powerMAX} for the increase in the output power of the heat engine using these values. Here we used a hot bath with temperature $\bar{n}_{\mathrm{ss}}=2$, and a cold bath with $\bar{n}_{\mathrm{ss}}=0.55$.} 
\label{fig:maxpower}
\end{figure*}
We can immediately deduce an improvement in the efficiency using the $\textsc{CC}$ thermal baths. Since the $\textsc{CC}$ thermal baths have a smaller shortcut-to-adiabaticity energy cost $V_{e}+V_{c}$ and a larger heat input $Q_{h}^{\textsc{cc}}$, then from \eqref{eq:efficiency} we must have
\begin{equation}
\eta^{\textsc{cc}} > \eta^{\textsc{i}}.\label{eq:effCC}
\end{equation}

For the $\textsc{CH}$ thermal baths, the coherence provides a reduction in the cost of the shortcut-to-adiabaticity protocol, however since $\Delta_{h}^{\textsc{ch}}<\Delta_{h}^{\textsc{i}}$, there is a reduction in the heat input to the system
\begin{equation}
Q_{h}^{\textsc{ch}} < Q_{h}^{\textsc{i}}.
\end{equation}
It can be shown (see Appendix~\ref{sec:AppEff} for details) that the reduction in heat into the system outweighs the benefit of coherence in reducing the cost of the shortcut-to-adiabaticity protocol, and we have (see FIG.~\ref{fig:efficiency})
\begin{equation}
\eta^{\textsc{ch}} < \eta^{\textsc{i}}.\label{eq:effCH}
\end{equation}

\subsection{Power}

The power generated from a single thermal cycle (of duration $t_{\text{cycle}}$) is the work per unit time
\begin{equation}
P = \frac{W-V_{e}-V_{c}}{t_{\text{cycle}}}.
\end{equation}
The work extracted is $W=\hbar\left(\omega_{h}-\omega_{c}\right)\left(\bar{n}_{h}-\bar{n}_{c}\right)$. Using \eqref{eq:nh-nc}, we can rewrite the power as
\begin{align}
P = &\;\frac{\hbar\left(\omega_{h}-\omega_{c}\right)}{t_{\text{cycle}}}\left(\frac{E_{h}}{\Delta_{h}}-\frac{E_{c}}{\Delta_{c}}\right)\frac{2\sinh\left(\frac{\Delta_{c}t_{c}}{2}\right)\sinh\left(\frac{\Delta_{h}t_{h}}{2}\right)}{\sinh\left(\frac{\Delta_{c}t_{c}+\Delta_{h}t_{h}}{2}\right)}\nonumber\\
&-\frac{V_{e}+V_{c}}{t_{\text{cycle}}}. \label{eq:power}
\end{align}
FIG.~\ref{fig:power} compares $P\left(t_{\text{cycle}}\right)$ for the different thermal baths. As we saw in the previous section, before time $t^{\prime}$ the cost $V_{e}+V_{c}$ is greater than the work extracted, and the engine becomes too costly to run. As $t_{h},t_{c}\rightarrow \infty$, the work approaches a constant value
\begin{equation}
\lim_{t_{h},t_{c}\rightarrow\infty} W = \hbar\left(\omega_{h}-\omega_{c}\right)\left(\frac{E_{h}}{\Delta_{h}}-\frac{E_{c}}{\Delta_{c}}\right),
\end{equation}
and $P\rightarrow 0$. Since $P$ is the product of transcendental functions, we cannot find an analytic expression for the cycle time $t_{\text{cycle}}$ at which $P$ is maximal. \\ 
To reiterate, the work $W$ is proportional to \mbox{$\bar{n}_{h}-\bar{n}_{c}$} which is monotonic in $\Delta_{h}$ and $\Delta_{c}$ (see Appendix~\ref{sec:AppWork}). Since $\Delta_{c}^{\textsc{cc}}>\Delta_{c}^{\textsc{I}}$, the work extracted from the \textsc{CC} baths in a cycle is greater than the incoherent (\textsc{I}) baths.
The \textsc{CC} baths extract more work, and have a smaller shortcut-to-adiabaticity energy cost, so we can immediately deduce an increase in power (FIG.~\ref{fig:power})
\begin{equation}
P^{\textsc{cc}}>P^{\textsc{i}}.\label{eq:powerCC}
\end{equation}
The \textsc{CH} baths also have a smaller energy cost, however $\Delta_{h}^\textsc{ch}<\Delta_{h}^{\textsc{i}}$, which implies that less work is extracted in a cycle from the \textsc{CH} baths than the incoherent (\textsc{I}) baths. The power of the \textsc{CH} baths need not be less than the incoherent baths (\textsc{I}), but in our simulations we find it typically is (see Appendix~\ref{sec:AppPower} for details). In FIG.~\ref{fig:power}, the decrease in work extracted in a single cycle is large enough such that there is a net decrease in the power $P^{\textsc{ch}}$ of the \textsc{CH} baths, i.e. $P^{\textsc{ch}}<P^{\textsc{i}}$.

\section{Optimising the Power}\label{sec:optimise}
\subsection{Using coherence}

Can we further improve the output power?
So far we have been looking at two extremes: using coherence to generate a hot bath (\textsc{CH}) and a cold bath (\textsc{CC}), with the other bath in each pair having no coherence (refer to FIG.~\ref{fig:bath_table}). But what about in-between these two extremes? We could compare these baths with others in which both hot and cold baths are created using coherence. We could look at the continuum in between these two extremes, parametrised by $\pi$:
\begin{align}
E_{h}^{\pi} = E_{c}^{\pi} &= \pi E_{h}^{\textsc{i}} + (1-\pi) E_{c}^{\textsc{i}}, \nonumber \\
\Delta_{h}^{\pi} &= \left(\pi E_{h}^{\textsc{i}} + (1-\pi) E_{c}^{\textsc{i}}\right)\frac{\Delta_{h}^{\textsc{i}}}{E_{h}^{\textsc{i}}}, \\
\Delta_{c}^{\pi} &= \left(\pi E_{h}^{\textsc{i}} + (1-\pi) E_{c}^{\textsc{i}}\right)\frac{\Delta_{c}^{\textsc{i}}}{E_{c}^{\textsc{i}}}, \nonumber
\end{align}
where $0\leq \pi\leq 1$. The limiting cases are those studied in the previous section: $\pi=0$ corresponds to the \textsc{CH} baths and $\pi=1$ corresponds to the \textsc{CC} baths. Along this continuum, the maximal work per cycle is achieved at $\pi=1$. In the previous section we mentioned numerous times (shown in Appendix~\ref{sec:AppWork}) that the work extracted in a single cycle is monotonically increasing with $\Delta_{h}$ and $\Delta_{c}$. Since $E_{h}^\textsc{i}>E_{c}^{\textsc{i}}$, we see that $\Delta_{h}^{\pi}$ and $\Delta_{c}^{\pi}$ are maximised when $\pi=1$. Using coherence to decrease the temperature of a bath increases the work extracted per cycle, and using coherence to increase the temperature of a bath decreases the work extracted per cycle. Thus it is expected to be that the optimal bath combination would be to use coherence to cool the cold bath, and to use no coherence in heating the hot bath. 

This reasoning suggests the way to maximise the work extracted per cycle using coherence is to create both the hot and cold baths from hotter baths, using as much coherence as possible to decrease their effective temperatures. However as we saw in section~\ref{sec:interaction}, if a system is cooled maximally, then the effective temperature $T_{\text{eff}}\leq \hbar\omega/\ln\left(2\right)$. Such a bath can be created using coherence to cool a thermal bath with temperature
\begin{equation}
T_{R}=\frac{\hbar\omega}{\beta_{\text{eff}}\hbar\omega-\ln\left(2\right)}.
\end{equation}
A bath with temperature $T_{\text{eff}}\geq \hbar\omega/\ln\left(2\right)$ can be created by using coherence to cool a bath of infinite temperature.

\subsection{Changing stroke times}

Hitherto, we have been assuming that for a given cycle time $t_{\text{cycle}}$, each stroke had equal duration $t_{\text{cycle}}/4$. The system interacts with the heat baths for a total time $t_{Q}=t_{h}+t_{c}$, and work is done on the system for a time $t_{W} = t_{\text{cycle}}-t_{Q}$. We can break down $t_{W}$ further, as $t_{W} = t_{W_{e}}+t_{W_{c}}$, where $t_{W_{e}}$ is the duration of the isentropic expansion, and $t_{W_{c}}$ is the duration of the isentropic compression.

We can improve the output power of the heat engine by altering all the relative times of the four cycle strokes. We parametrise this using three variables.

First, we can alter the relative durations of the heat and work strokes within a single cycle. We parametrise the total cycle time $t_{\text{cycle}}$ as follows
\begin{equation}
\begin{matrix}
t_{Q} = p \times t_{\text{cycle}} \\
t_{W} = (1-p) \times t_{\text{cycle}}
\end{matrix}.
\label{eq:Ptcycle}
\end{equation}

Increasing the proportion $p$ of $t_{\text{cycle}}$ in which the system interacts with the heat bath causes more heat to flow in and out of the system, allowing more work to be extracted per cycle, at the price of a larger energy cost in running the shortcut-to-adiabaticity protocol.
We want to find the value of $p$ in which the peak power is maximised.
We cannot find an analytic expression for an optimal value for $p$, so we proceed numerically.

We can further parametrise $t_{Q}$ and $t_{W}$,
\begin{equation}
\begin{matrix}
t_{h} = q\times t_{Q} & t_{W_{e}} = r\times t_{W} \\
t_{c} = (1-q)\times t_{Q} & t_{W_{c}} = (1-r)\times t_{W} 
\end{matrix}.\label{eq:paramqr}
\end{equation}
Increasing the proportion $q$ of $t_{Q}$ which the system interacts with the hot bath allows more heat to enter the system, increasing the work that can be extracted. However less heat is transferred to the cold bath, so more work is required during the compression stroke. 
The costs of expansion ($V_{e}$) and compression ($V_{c}$) differ due to the different initial average photon numbers (see FIG.~\ref{fig:expand_compress}). Increasing $r$ reduces the cost of the isentropic expansion strokes; however, it increases the cost of the isentropic compression strokes.

\begin{figure}[t]
\includegraphics[scale=0.4]{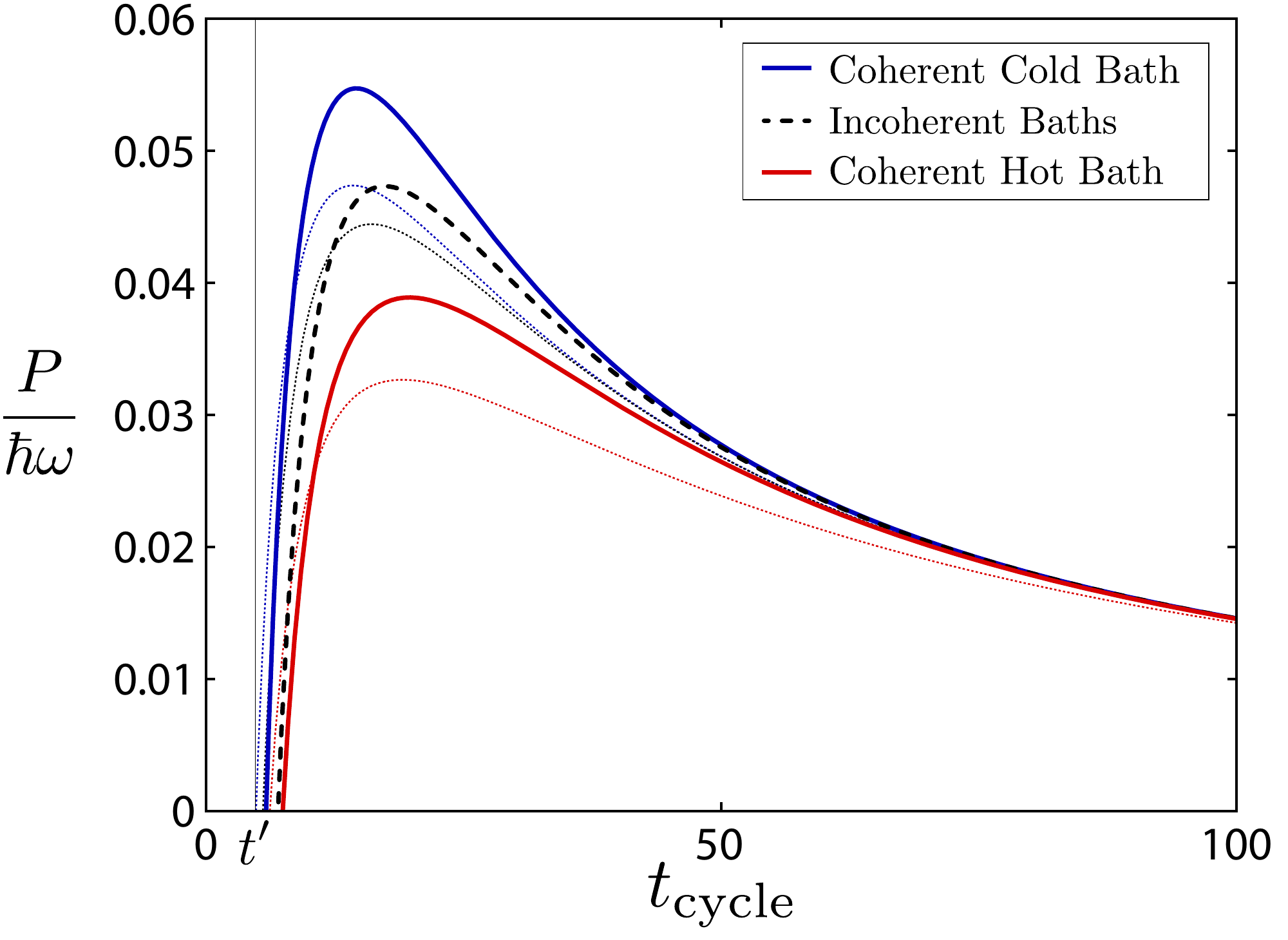}
\caption{The output power can be improved by altering the relative times of each stroke of the Otto cycle. See FIG.~\ref{fig:maxpower} for the numerical search for the optimal values of $p,q,r$ for the \textsc{CC} baths. Power with $p=0.58$, $q=0.69$ and $r=0.59$ in thick lines, showing noticeable enhancement. Dotted lines show $p=q=r=0.5$ (where the duration of all strokes is $t_{\text{cycle}}/4$, i.e. FIG.~\ref{fig:power}). The power is scaled by $1/\hbar\omega$, where $\omega= 10^{14}$Hz. Recall that $t_{\text{cycle}}$ is in terms of the dimensionless time parameter \eqref{eq:me}. The three pairs of thermal baths are described in FIG.~\ref{fig:bath_table}. Here we used a hot bath with temperature $\bar{n}_{\mathrm{ss}}=2$, and a cold bath with $\bar{n}_{\mathrm{ss}}=0.55$.}
\label{fig:powerMAX}
\end{figure}
We want to find the values of $p,q,r$ such that the peak power is maximised. That is, we want to maximise the following function
\begin{equation}
\text{Max Power}\left(p,q,r\right) = \max_{t_{\text{cycle}}} P\left(t_{\text{cycle}},p,q,r\right).
\label{eq:maxpqr}
\end{equation}
In FIG.~\ref{fig:maxpower}, three cross sections of this function are shown, one for each variable. We see the values which maximise the power are different for each of the baths. We numerically found values of $p,q,r$ which maximised the Max Power for the \textsc{CC} baths with arbitrarily chosen temperatures. In this case we found a bias towards increasing the duration of the heat strokes over the work strokes, and in particular a higher proportion of the interaction time $t_{Q}$ to be with the hot bath. The power is also improved by slightly slowing the expansion strokes.
In FIG.~\ref{fig:powerMAX} the improvement in the power curve with these values is shown.

\section{Conclusion}

Coherence in the energy ground space of the atoms used to emulate a thermal bath does not change the average energy of the atoms, but changes the interaction of the atoms with the system; specifically the rate $G$ at which the system loses photons to the incoming atoms. This has two effects. First, it changes the thermalisation temperature of the system. So coherence can be used to create a temperature difference in two baths which previously were in equilibrium. Secondly, it changes the rate of thermalisation $\Delta$. These effects are inversely correlated, in that decreasing the thermalisation temperature causes an increase in the rate of thermalisation, and vice versa. In this paper we have shown how the coherence can be used to increase (or decrease) the power and efficiency of a heat engine. We used coherence to create a temperature difference between two bath systems in equilibrium; creating both a cold bath, and a hot bath. In both cases, the coherence reduced the cost of implementing the shortcut-to-adiabaticity protocol. Using coherence to create the cold bath improved the power of the engine, because it increased the thermalisation rate $\Delta$, allowing more heat transfer into the system in a cycle, and more work to be extracted. The efficiency was also improved, because the increased heat into the system reduced the effect of the cost of the shortcut-to-adiabaticity protocol. Conversely, creating a hot bath requires the thermalisation rate $\Delta$ to be reduced, which can in turn reduce the output power of the engine, and always reduces the efficiency. This demonstrates a more nuanced and subtle perspective of the benefit of coherence in the finite-time regime, it is a quantity which requires careful optimisation.
The output power can be further improved by using the maximum amount of coherence in cooling the heat baths, also by changing the relative durations of the four strokes in the thermal cycle.

This work can be extended by considering a quantum trajectory unravelling of the master equation \eqref{eq:me}. This would involve continuously monitoring the photon number of the system, and using that information to reduce the duration of the heat strokes. Of course the memory which stores the measurement results would need to be erased to close the cycle.
Considering atoms with a larger number of degenerate ground states, might allow more efficient energy transfer between the bath and system.
Furthermore, correlated states between the atoms would allow the study of more general quantum resources, extending beyond Markovian interactions.

\begin{acknowledgements}
T. Guff would like to thank J. D. Cresser for his many useful conversations.
This research was funded in part by the Australian Research Council Centre of Excellence for Engineered Quantum Systems (Project number CE110001013, CE170100009). 
\end{acknowledgements}

\appendix
\section{Steady Cycle}\label{sec:Steady_cycle}
After driving the heat engine through $\delta$ Otto cycles, using the shortcut-to-adiabaticity techniques during the work strokes, the average photon number is \eqref{eq:aftercycle}
\begin{align}
    \bar{n}_{\delta} = &\left(\left(\bar{n}_{\delta-1}-\frac{E_{h}}{\Delta_{h}}\right)e^{-\Delta_{h}t_{h}}+\frac{E_{h}}{\Delta_{h}}-\frac{E_{c}}{\Delta_{c}}\right) \nonumber\\
    &\times e^{-\Delta_{c}t_{c}} + \frac{E_{c}}{\Delta_{c}}.
\end{align}
where $\bar{n}_{\delta-1}$ is the average photon number at the end of $\delta-1$ cycles. If we recursively expand this using \eqref{eq:nt} until we reach the initial photon number $\bar{n}_{0}$, we find (with some algebra)
\begin{align}
\bar{n}_{\delta}= &\;\bar{n}_{0} e^{-\delta\Delta_{c} t_{c}} e^{-\delta\Delta_{h} t_{h}}\nonumber\\
&+\left[\frac{E_{h}}{\Delta_{h}}e^{-\Delta_{c} t_{c}}\left(1-e^{-\Delta_{h} t_{h}}\right)+\frac{E_{c}}{\Delta_{c}}\left(1 -e^{-\Delta_{c} t_{c}}\right)\right]\nonumber \\
&\times \left[\underbrace{e^{-\Delta_{c} t_{c}}e^{-\Delta_{h} t_{h}} (\dots (e^{-\Delta_{c} t_{c}} e^{-\Delta_{h} t_{h}}}_{\delta-1\text{ terms}}+\underbrace{1 )\dots )+1}_{\delta-1\text{ terms}}\right].
\label{eq:afterpcycles}
\end{align}
From \eqref{eq:steadycycle} we can identify
\begin{align}
\frac{E_{h}}{\Delta_{h}}e^{-\Delta_{c} t_{c}}&\left(1-e^{-\Delta_{h} t_{h}}\right)+\frac{E_{c}}{\Delta_{c}}\left(1 -e^{-\Delta_{c} t_{c}}\right) \nonumber \\
=-\bar{n}_{sc}&\left(e^{-\Delta_{c}t_{c}}e^{-\Delta_{h}t_{h}}-1\right).
\end{align}
Further we can identify the geometric series
\begin{align}
&\underbrace{e^{-\Delta_{c} t_{c}}e^{-\Delta_{h} t_{h}} (\dots (e^{-\Delta_{c} t_{c}} e^{-\Delta_{h} t_{h}}}_{\delta-1\text{ terms}}+\underbrace{1 )\dots )+1}_{\delta-1\text{ terms}}\nonumber \\
&= \sum_{j=0}^{\delta-1} e^{-j\Delta_{c}t_{c}}e^{-j\Delta_{h}t_{h}} = \frac{e^{-\delta\Delta_{c}t_{c}}e^{-\delta\Delta_{h}t_{h}}-1}{e^{-\Delta_{c}t_{c}}e^{-\Delta_{h}t_{h}}-1}.
\end{align}
We combine these to rewrite \eqref{eq:afterpcycles} as
\begin{equation}
\bar{n}_{\delta} = e^{-\delta\Delta_{c}t_{c}}e^{-\delta\Delta_{h}t_{h}}\left(\bar{n}_{0}-\bar{n}_{sc}\right) + \bar{n}_{sc}.
\end{equation}

\section{Derivatives}\label{sec:derivatives}

\subsection{Work}\label{sec:AppWork}

The work is proportional to  the change in the average photon number, $\bar{n}_{h}-\bar{n}_{c}$, which is given by \eqref{eq:nh-nc}. With the bath temperatures held constant, this function is monotonically increasing in $\Delta_{h}$ and $\Delta_{c}$. We see this by calculating derivatives,
\begin{subequations}
\begin{align}
\left.\frac{d \left(\bar{n}_{h}-\bar{n}_{c}\right)}{d\Delta_{h}}\right|_{\frac{E_{h}}{\Delta_{h}}} &= \alpha_{h}\left(\frac{E_{h}}{\Delta_{h}}-\frac{E_{c}}{\Delta_{c}}\right), \label{eq:dnh}\\
\left.\frac{d \left(\bar{n}_{h}-\bar{n}_{c}\right)}{d\Delta_{c}}\right|_{\frac{E_{c}}{\Delta_{c}}} &= \alpha_{c}\left(\frac{E_{h}}{\Delta_{h}}-\frac{E_{c}}{\Delta_{c}}\right), \label{eq:dnc}
\end{align} \label{eq:dn}
\end{subequations}
where
\begin{subequations}
\begin{align}
\alpha_{h} &= \frac{t_{h}}{2}\frac{\cosh\left(\Delta_{c}t_{c}\right)-1}{\cosh\left( \Delta_{c}t_{c}+\Delta_{h}t_{h}\right)-1} > 0, \\
\alpha_{c} &= \frac{t_{c}}{2}\frac{\cosh\left(\Delta_{h}t_{h}\right)-1}{\cosh\left(\Delta_{c}t_{c}+\Delta_{h}t_{h}\right)-1} > 0.
\end{align}
\end{subequations}
So both derivatives in \eqref{eq:dn} are both positive. Therefore more work is extracted per cycle by increasing $\Delta_{h}$ and $\Delta_{c}$.

\subsection{Cost}\label{sec:AppCost}
The cost of implementing the shortcut-to-adiabaticity protocol depends on the thermal baths in that it is proportional to the initial average photon number \eqref{eq:cost}. At the beginning of the compression strokes, the initial photon number is given by the steady cycle initial photon number $\bar{n}_{sc}$ \eqref{eq:steadycycle}. By calculating the derivatives of this function, holding the temperature of the baths constant, we find
\begin{subequations}
\begin{align}
\left.\frac{d \bar{n}_{sc}}{d\Delta_{h}}\right|_{\frac{E_{h}}{\Delta_{h}}} &= \gamma_{h}\left(\frac{E_{h}}{\Delta_{h}}-\frac{E_{c}}{\Delta_{c}}\right), \\
\left.\frac{d \bar{n}_{sc}}{d\Delta_{c}}\right|_{\frac{E_{c}}{\Delta_{c}}} &= -\gamma_{c}\left(\frac{E_{h}}{\Delta_{h}}-\frac{E_{c}}{\Delta_{c}}\right),
\end{align}
\end{subequations}
where
\begin{subequations}
\begin{align}
\gamma_{h} &= \frac{t_{h}}{2}\frac{e^{\frac{-\Delta_{c}t_{c}}{2}}\sinh\left(\frac{ \Delta_{c}t_{c}}{2}\right)}{\sinh^{2}\left(\frac{\Delta_{c}t_{c}+\Delta_{h}t_{h}}{2}\right)} > 0, \\
\gamma_{c} &= \frac{t_{c}}{2}\frac{e^{\frac{\Delta_{h}t_{h}}{2}}\sinh\left(\frac{\Delta_{h}t_{h}}{2}\right)}{\sinh^{2}\left(\frac{\Delta_{c}t_{c}+\Delta_{h}t_{h}}{2}\right)} > 0,
\end{align}
\end{subequations}
so $\bar{n}_{sc}$ is monotonically increasing in $\Delta_{h}$, but monotonically decreasing in $\Delta_{c}$. The expansion strokes begin after the system has interacted with the hot bath for a time $t_{h}$. That is, the system begins with $\bar{n}\left(0\right)=\bar{n}_{sc}$, and then evolves under \eqref{eq:nt} for time $t_{h}$, interacting with the hot bath, after which the average photon number $\bar{n}_{h}$ is
\begin{equation}
\bar{n}_{h} = \frac{\frac{E_{h}}{\Delta_{h}}e^{\frac{\Delta_{c}t_{c}}{2}}\sinh\left(\frac{\Delta_{h}t_{h}}{2}\right)+\frac{E_{c}}{\Delta_{c}}e^{\frac{-\Delta_{h}t_{h}}{2}}\sinh\left(\frac{\Delta_{c}t_{c}}{2}\right)}{\sinh\left(\frac{\Delta_{h}t_{h}+\Delta_{c}t_{c}}{2}\right)}. \label{eq:nh}
\end{equation}
If the interaction time with the hot bath $t_{h}$ is large, then the the average photon number is the thermalisation photon number of the hot bath
\begin{equation}
\lim_{t_{h}\rightarrow\infty} \bar{n}_{h} = \frac{E_{h}}{\Delta_{h}}.
\end{equation}
Once again taking derivatives we find
\begin{subequations}
\begin{align}
\left.\frac{d \bar{n}_{h}}{d\Delta_{h}}\right|_{\frac{E_{h}}{\Delta_{h}}} &= \xi_{h}\left(\frac{E_{h}}{\Delta_{h}}-\frac{E_{c}}{\Delta_{c}}\right), \\
\left.\frac{d \bar{n}_{h}}{d\Delta_{c}}\right|_{\frac{E_{c}}{\Delta_{c}}} &= -\xi_{c}\left(\frac{E_{h}}{\Delta_{h}}-\frac{E_{c}}{\Delta_{c}}\right),
\end{align}
\end{subequations}
where
\begin{subequations}
\begin{align}
\xi_{h} &= \frac{t_{h}}{2}\frac{e^{\frac{\Delta_{c}t_{c}}{2}}\sinh\left(\frac{\Delta_{c}t_{c}}{2}\right)}{\sinh^{2}\left(\frac{\Delta_{c}t_{c}+\Delta_{h}t_{h}}{2}\right)} > 0, \\
\xi_{c} &= \frac{t_{c}}{2}\frac{e^{\frac{-\Delta_{h}t_{h}}{2}}\sinh\left(\frac{\Delta_{h}t_{h}}{2}\right)}{\sinh^{2}\left(\frac{\Delta_{c}t_{c}+\Delta_{h}t_{h}}{2}\right)} > 0,
\end{align}
\end{subequations}
and likewise, $\bar{n}_{h}$ is monotonically increasing in $\Delta_{h}$, but monotonically decreasing in $\Delta_{c}$.

\subsection{Efficiency}\label{sec:AppEff}

The efficiency \eqref{eq:efficiencywithcost} is reduced from maximum by the cost of implementing the shortcut-to-adiabaticity protocol, over the heat input $Q_{h}$ into the system. The work costs $V_{e}$ and $V_{c}$ are proportional to the average photon number of the system at the beginning of the expansion and compression. Therefore we can write $V_{e} = \bar{n}_{h}I_{e}$ and $V_{c}=\bar{n}_{c}I_{c}$, where (as can be seen in \eqref{eq:cost}), $I_{e}$ and $I_{c}$ are integrals which are independent of the choice of bath.

The \textsc{CC} baths have a smaller energy cost, and a larger heat input per cycle than the incoherent (\textsc{I}) baths: therefore the efficiency of the \textsc{CC} baths is always greater. However, the \textsc{CH} baths have a smaller shortcut-to-adiabaticity energy cost $V_{e}^{\textsc{ch}}+V_{c}^{\textsc{ch}}$ compared with the incoherent (\textsc{I}) baths, but also a smaller heat input $Q_{h}^{\textsc{ch}}$.
To see the effect on the efficiency in using the $\textsc{CH}$ baths, we need to compare
\begin{equation}
\frac{\bar{n}_{h}^{\textsc{i}}I_{e}+\bar{n}_{c}^{\textsc{i}}I_{c}}{\bar{n}_{h}^{\textsc{i}}-\bar{n}_{c}^{\textsc{i}}}
\text{  with  }
\frac{\bar{n}_{h}^{\textsc{ch}}I_{e}+\bar{n}_{c}^{\textsc{ch}}I_{c}}{\bar{n}_{h}^{\textsc{ch}}-\bar{n}_{c}^{\textsc{ch}}}.
\end{equation}
Using \eqref{eq:steadycycle}, \eqref{eq:nh-nc}, and \eqref{eq:nh}, we have
\begin{subequations}
\begin{align}
\zeta_{h} &= \frac{\bar{n}_{h}}{\bar{n}_{h}-\bar{n}_{c}} \nonumber \\
&= \frac{1}{2}+\frac{1}{2}\left(\frac{E_{h}}{\Delta_{h}}-\frac{E_{c}}{\Delta_{c}}\right)^{-1}\nonumber \\
&\quad\times
\left(\frac{E_{h}}{\Delta_{h}}\coth\left(\frac{\Delta_{c}t_{c}}{2}\right) - \frac{E_{c}}{\Delta_{c}}\coth\left(\frac{\Delta_{h}t_{h}}{2}\right)\right), \\ 
\zeta_{c} &= \frac{\bar{n}_{c}}{\bar{n}_{h}-\bar{n}_{c}} \nonumber \\
&= -\frac{1}{2}+\frac{1}{2}\left(\frac{E_{h}}{\Delta_{h}}-\frac{E_{c}}{\Delta_{c}}\right)^{-1}\nonumber \\
&\quad\times
\left(\frac{E_{h}}{\Delta_{h}}\coth\left(\frac{\Delta_{c}t_{c}}{2}\right) - \frac{E_{c}}{\Delta_{c}}\coth\left(\frac{\Delta_{h}t_{h}}{2}\right)\right).
\end{align}
\end{subequations}
The difference between $\zeta_{h}$ and $\zeta_{c}$ is constant.
We want the derivative with respect to $\Delta_{h}$, keeping the temperature of the hot bath constant.
\begin{align}
&\left.\frac{d\zeta_{h}}{d\Delta_{h}}\right|_{\frac{E_{h}}{\Delta_{h}}}=\left.\frac{d\zeta_{c}}{d\Delta_{h}}\right|_{\frac{E_{h}}{\Delta_{h}}} \nonumber\\
&=\frac{-t_{h}}{4}\frac{E_{c}}{\Delta_{c}}\left(\frac{E_{h}}{\Delta_{h}}-\frac{E_{c}}{\Delta_{c}}\right)^{-1}\frac{1}{\sinh\left(\frac{\Delta_{h}t_{h}}{2}\right)} < 0,
\end{align}
so $\zeta_{h}$ and $\zeta_{c}$ are monotonically decreasing with $\Delta_{h}$.
Changing from the $\textsc{I}$ thermal baths to the $\textsc{CH}$ thermal baths involves decreasing $\Delta_{h}$, which increases $\zeta_{h}$ and $\zeta_{c}$. Thus we have
\begin{equation}
\frac{\bar{n}_{h}^{\textsc{i}}I_{e}+\bar{n}_{c}^{\textsc{i}}I_{c}}{\bar{n}_{h}^{\textsc{i}}-\bar{n}_{c}^{\textsc{i}}} <\frac{\bar{n}_{h}^{\textsc{ch}}I_{e}+\bar{n}_{c}^{\textsc{ch}}I_{c}}{\bar{n}_{h}^{\textsc{ch}}-\bar{n}_{c}^{\textsc{ch}}}.
\end{equation}
So
\begin{equation}
\frac{V_{e}^{\textsc{i}}+V_{c}^{\textsc{i}}}{Q_{h}^{\textsc{i}}}<\frac{V_{e}^{\textsc{ch}}+V_{c}^{\textsc{ch}}}{Q_{h}^{\textsc{ch}}},
\end{equation}
and hence
\begin{equation}
\eta^{\textsc{ch}}<\eta^{\textsc{i}}.
\end{equation}

\subsection{Power}\label{sec:AppPower}

The \textsc{CC} baths extract more work per cycle than the incoherent (\textsc{I}) baths, and also have a smaller shortcut-to-adiabaticity energy cost. Hence the output power of the \textsc{CC} is always larger than the incoherent (\textsc{I}) baths.\\
However, the work extracted in a single cycle using the \textsc{CH} baths is \emph{less} than the \textsc{I} baths, and the use of coherence means there is a smaller shortcut-to-adiabaticity cost, $V_{e}^{\textsc{ch}}+V_{c}^{\textsc{ch}}<V_{e}^{\textsc{i}}+V_{c}^{\textsc{i}}$.\\
The output power of the thermal engine is given by \eqref{eq:power}
\begin{align*}
P = &\;\frac{\hbar\left(\omega_{h}-\omega_{c}\right)}{t_{\text{cycle}}}\left(\frac{E_{h}}{\Delta_{h}}-\frac{E_{c}}{\Delta_{c}}\right)\frac{2\sinh\left(\frac{\Delta_{c}t_{c}}{2}\right)\sinh\left(\frac{\Delta_{h}t_{h}}{2}\right)}{\sinh\left(\frac{\Delta_{c}t_{c}+\Delta_{h}t_{h}}{2}\right)}\nonumber\\
&-\frac{V_{e}+V_{c}}{t_{\text{cycle}}}. 
\end{align*}
As in the previous section we can rewrite the energy cost
\begin{equation}
V_{e}+V_{c} = \bar{n}_{h}I_{e}+\bar{n}_{c}I_{c},
\end{equation}
where $\bar{n}_{h}$ is given by \eqref{eq:nh} and $\bar{n}_{c}$ is given by \eqref{eq:steadycycle}. We want to know whether the power of the \textsc{CH} baths is strictly less than the incoherent (\textsc{I}) baths. 
So we want the derivative of $P$ with respect to $\Delta_{h}$, keeping temperature constant.
\begin{align}
\left.\frac{dP}{d\Delta_{h}}\right|_{\frac{E_{h}}{\Delta_{h}}} &= \frac{t_{h}}{t_{\text{cycle}}}\left(\frac{E_{h}}{\Delta_{h}}-\frac{E_{c}}{\Delta_{c}}\right)\frac{e^{\Delta_{c}t_{c}}-1}{\left(e^{\Delta_{c}t_{c}+\Delta_{h}t_{h}}\right)^{2}}\nonumber \\
&\quad\times\left(\hbar\left(\omega_{h}-\omega_{c}\right)\left(e^{\Delta_{c}t_{c}}-1\right)-I_{e}e^{\Delta_{c}t_{c}}-I_{c}\right).
\end{align}
The power $P$ is monotonically increasing in $\Delta_{h}$ if
\begin{equation}
e^{\Delta_{c}t_{c}}>\frac{\hbar\left(\omega_{h}-\omega_{c}\right)+I_{c}}{\hbar\left(\omega_{h}-\omega_{c}\right)-I_{e}}.
\end{equation}
If the difference in system frequencies $\hbar\left(\omega_{h}-\omega_{c}\right)$ is large compared with the energy costs $I_{e},I_{c}$, then $P$ will be monotonically increasing, and the power $P^{\textsc{ch}}$ of the \textsc{CH} baths will be less than that of the incoherent (\textsc{I}) baths. However, if $\hbar\left(\omega_{h}-\omega_{c}\right)\approx I_{e}$ and $\hbar\left(\omega_{h}-\omega_{c}\right) > I_{e}$, then $P$ will be monotonically decreasing for all but large values of $\Delta_{c}t_{c}$, and the power \textsc{CH} baths will exceed that of the incoherent (\textsc{I}) baths.

\bibliography{references}

\begin{thebibliography}{50}%
\makeatletter
\providecommand \@ifxundefined [1]{%
 \@ifx{#1\undefined}
}%
\providecommand \@ifnum [1]{%
 \ifnum #1\expandafter \@firstoftwo
 \else \expandafter \@secondoftwo
 \fi
}%
\providecommand \@ifx [1]{%
 \ifx #1\expandafter \@firstoftwo
 \else \expandafter \@secondoftwo
 \fi
}%
\providecommand \natexlab [1]{#1}%
\providecommand \enquote  [1]{``#1''}%
\providecommand \bibnamefont  [1]{#1}%
\providecommand \bibfnamefont [1]{#1}%
\providecommand \citenamefont [1]{#1}%
\providecommand \href@noop [0]{\@secondoftwo}%
\providecommand \href [0]{\begingroup \@sanitize@url \@href}%
\providecommand \@href[1]{\@@startlink{#1}\@@href}%
\providecommand \@@href[1]{\endgroup#1\@@endlink}%
\providecommand \@sanitize@url [0]{\catcode `\\12\catcode `\$12\catcode
  `\&12\catcode `\#12\catcode `\^12\catcode `\_12\catcode `\%12\relax}%
\providecommand \@@startlink[1]{}%
\providecommand \@@endlink[0]{}%
\providecommand \url  [0]{\begingroup\@sanitize@url \@url }%
\providecommand \@url [1]{\endgroup\@href {#1}{\urlprefix }}%
\providecommand \urlprefix  [0]{URL }%
\providecommand \Eprint [0]{\href }%
\providecommand \doibase [0]{http://dx.doi.org/}%
\providecommand \selectlanguage [0]{\@gobble}%
\providecommand \bibinfo  [0]{\@secondoftwo}%
\providecommand \bibfield  [0]{\@secondoftwo}%
\providecommand \translation [1]{[#1]}%
\providecommand \BibitemOpen [0]{}%
\providecommand \bibitemStop [0]{}%
\providecommand \bibitemNoStop [0]{.\EOS\space}%
\providecommand \EOS [0]{\spacefactor3000\relax}%
\providecommand \BibitemShut  [1]{\csname bibitem#1\endcsname}%
\let\auto@bib@innerbib\@empty
\bibitem [{\citenamefont {Goold}\ \emph {et~al.}(2016)\citenamefont {Goold},
  \citenamefont {Huber}, \citenamefont {Riera}, \citenamefont {Rio},\ and\
  \citenamefont {Skrzypczyk}}]{2016Goold143001}%
  \BibitemOpen
  \bibfield  {author} {\bibinfo {author} {\bibfnamefont {J.}~\bibnamefont
  {Goold}}, \bibinfo {author} {\bibfnamefont {M.}~\bibnamefont {Huber}},
  \bibinfo {author} {\bibfnamefont {A.}~\bibnamefont {Riera}}, \bibinfo
  {author} {\bibfnamefont {L.~d.}\ \bibnamefont {Rio}}, \ and\ \bibinfo
  {author} {\bibfnamefont {P.}~\bibnamefont {Skrzypczyk}},\ }\href {\doibase
  10.1088/1751-8113/49/14/143001} {\bibfield  {journal} {\bibinfo  {journal}
  {J. Phys. A}\ }\textbf {\bibinfo {volume} {49}},\ \bibinfo {pages} {143001}
  (\bibinfo {year} {2016})}\BibitemShut {NoStop}%
\bibitem [{\citenamefont {Vinjanampathy}\ and\ \citenamefont
  {Anders}(2016)}]{2016Vinjanampathy545}%
  \BibitemOpen
  \bibfield  {author} {\bibinfo {author} {\bibfnamefont {S.}~\bibnamefont
  {Vinjanampathy}}\ and\ \bibinfo {author} {\bibfnamefont {J.}~\bibnamefont
  {Anders}},\ }\href {\doibase 10.1080/00107514.2016.1201896} {\bibfield
  {journal} {\bibinfo  {journal} {Contemp. Phys.}\ }\textbf {\bibinfo {volume}
  {57}},\ \bibinfo {pages} {545} (\bibinfo {year} {2016})}\BibitemShut
  {NoStop}%
\bibitem [{\citenamefont {Erez}(2012)}]{Erez2012}%
  \BibitemOpen
  \bibfield  {author} {\bibinfo {author} {\bibfnamefont {N.}~\bibnamefont
  {Erez}},\ }\href {\doibase 10.1088/0031-8949/2012/T151/014028} {\bibfield
  {journal} {\bibinfo  {journal} {Phys. Script.}\ }\textbf {\bibinfo {volume}
  {T151}},\ \bibinfo {pages} {014028} (\bibinfo {year} {2012})}\BibitemShut
  {NoStop}%
\bibitem [{\citenamefont {Kammerlander}\ and\ \citenamefont
  {Anders}(2016)}]{Kammerlander2016}%
  \BibitemOpen
  \bibfield  {author} {\bibinfo {author} {\bibfnamefont {P.}~\bibnamefont
  {Kammerlander}}\ and\ \bibinfo {author} {\bibfnamefont {J.}~\bibnamefont
  {Anders}},\ }\href {\doibase 10.1038/srep22174} {\bibfield  {journal}
  {\bibinfo  {journal} {Sci. Rep.}\ }\textbf {\bibinfo {volume} {6}},\ \bibinfo
  {pages} {22174} (\bibinfo {year} {2016})}\BibitemShut {NoStop}%
\bibitem [{\citenamefont {Jacobs}(2012)}]{Jacobs2012}%
  \BibitemOpen
  \bibfield  {author} {\bibinfo {author} {\bibfnamefont {K.}~\bibnamefont
  {Jacobs}},\ }\href {\doibase 10.1103/PhysRevE.86.040106} {\bibfield
  {journal} {\bibinfo  {journal} {Phys. Rev. E}\ }\textbf {\bibinfo {volume}
  {86}},\ \bibinfo {pages} {040106} (\bibinfo {year} {2012})}\BibitemShut
  {NoStop}%
\bibitem [{\citenamefont {Ding}\ \emph {et~al.}(2018)\citenamefont {Ding},
  \citenamefont {Yi}, \citenamefont {Kim},\ and\ \citenamefont
  {Talkner}}]{Ding2018}%
  \BibitemOpen
  \bibfield  {author} {\bibinfo {author} {\bibfnamefont {X.}~\bibnamefont
  {Ding}}, \bibinfo {author} {\bibfnamefont {J.}~\bibnamefont {Yi}}, \bibinfo
  {author} {\bibfnamefont {Y.~W.}\ \bibnamefont {Kim}}, \ and\ \bibinfo
  {author} {\bibfnamefont {P.}~\bibnamefont {Talkner}},\ }\href {\doibase
  10.1103/PhysRevE.98.042122} {\bibfield  {journal} {\bibinfo  {journal} {Phys.
  Rev. E}\ }\textbf {\bibinfo {volume} {98}},\ \bibinfo {pages} {042122}
  (\bibinfo {year} {2018})}\BibitemShut {NoStop}%
\bibitem [{\citenamefont {Yi}\ \emph {et~al.}(2017)\citenamefont {Yi},
  \citenamefont {Talkner},\ and\ \citenamefont {Kim}}]{Yi2017}%
  \BibitemOpen
  \bibfield  {author} {\bibinfo {author} {\bibfnamefont {J.}~\bibnamefont
  {Yi}}, \bibinfo {author} {\bibfnamefont {P.}~\bibnamefont {Talkner}}, \ and\
  \bibinfo {author} {\bibfnamefont {Y.~W.}\ \bibnamefont {Kim}},\ }\href
  {\doibase 10.1103/PhysRevE.96.022108} {\bibfield  {journal} {\bibinfo
  {journal} {Phys. Rev. E}\ }\textbf {\bibinfo {volume} {96}},\ \bibinfo
  {pages} {022108} (\bibinfo {year} {2017})}\BibitemShut {NoStop}%
\bibitem [{\citenamefont {Funo}\ \emph {et~al.}(2013)\citenamefont {Funo},
  \citenamefont {Watanabe},\ and\ \citenamefont {Ueda}}]{Funo2013}%
  \BibitemOpen
  \bibfield  {author} {\bibinfo {author} {\bibfnamefont {K.}~\bibnamefont
  {Funo}}, \bibinfo {author} {\bibfnamefont {Y.}~\bibnamefont {Watanabe}}, \
  and\ \bibinfo {author} {\bibfnamefont {M.}~\bibnamefont {Ueda}},\ }\href
  {\doibase 10.1103/PhysRevA.88.052319} {\bibfield  {journal} {\bibinfo
  {journal} {Phys. Rev. A}\ }\textbf {\bibinfo {volume} {88}},\ \bibinfo
  {pages} {052319} (\bibinfo {year} {2013})}\BibitemShut {NoStop}%
\bibitem [{\citenamefont {Maruyama}\ \emph {et~al.}(2009)\citenamefont
  {Maruyama}, \citenamefont {Nori},\ and\ \citenamefont
  {Vedral}}]{Maruyama2009}%
  \BibitemOpen
  \bibfield  {author} {\bibinfo {author} {\bibfnamefont {K.}~\bibnamefont
  {Maruyama}}, \bibinfo {author} {\bibfnamefont {F.}~\bibnamefont {Nori}}, \
  and\ \bibinfo {author} {\bibfnamefont {V.}~\bibnamefont {Vedral}},\ }\href
  {\doibase 10.1103/RevModPhys.81.1} {\bibfield  {journal} {\bibinfo  {journal}
  {Rev. Mod. Phys.}\ }\textbf {\bibinfo {volume} {81}},\ \bibinfo {pages} {1}
  (\bibinfo {year} {2009})}\BibitemShut {NoStop}%
\bibitem [{\citenamefont {Elouard}\ \emph {et~al.}(2017)\citenamefont
  {Elouard}, \citenamefont {Herrera-Mart{\'{i}}}, \citenamefont {Huard},\ and\
  \citenamefont {Auff{\`{e}}ves}}]{Elouard2017}%
  \BibitemOpen
  \bibfield  {author} {\bibinfo {author} {\bibfnamefont {C.}~\bibnamefont
  {Elouard}}, \bibinfo {author} {\bibfnamefont {D.}~\bibnamefont
  {Herrera-Mart{\'{i}}}}, \bibinfo {author} {\bibfnamefont {B.}~\bibnamefont
  {Huard}}, \ and\ \bibinfo {author} {\bibfnamefont {A.}~\bibnamefont
  {Auff{\`{e}}ves}},\ }\href {\doibase 10.1103/PhysRevLett.118.260603}
  {\bibfield  {journal} {\bibinfo  {journal} {Phys. Rev. Lett.}\ }\textbf
  {\bibinfo {volume} {118}},\ \bibinfo {pages} {260603} (\bibinfo {year}
  {2017})}\BibitemShut {NoStop}%
\bibitem [{\citenamefont {Elouard}\ and\ \citenamefont
  {Jordan}(2018)}]{Elouard2018}%
  \BibitemOpen
  \bibfield  {author} {\bibinfo {author} {\bibfnamefont {C.}~\bibnamefont
  {Elouard}}\ and\ \bibinfo {author} {\bibfnamefont {A.~N.}\ \bibnamefont
  {Jordan}},\ }\href {\doibase 10.1103/PhysRevLett.120.260601} {\bibfield
  {journal} {\bibinfo  {journal} {Phys. Rev. Lett.}\ }\textbf {\bibinfo
  {volume} {120}},\ \bibinfo {pages} {260601} (\bibinfo {year}
  {2018})}\BibitemShut {NoStop}%
\bibitem [{\citenamefont {Hasegawa}\ \emph {et~al.}(2010)\citenamefont
  {Hasegawa}, \citenamefont {Ishikawa}, \citenamefont {Takara},\ and\
  \citenamefont {Driebe}}]{Hasegawa2010}%
  \BibitemOpen
  \bibfield  {author} {\bibinfo {author} {\bibfnamefont {H.-H.}\ \bibnamefont
  {Hasegawa}}, \bibinfo {author} {\bibfnamefont {J.}~\bibnamefont {Ishikawa}},
  \bibinfo {author} {\bibfnamefont {K.}~\bibnamefont {Takara}}, \ and\ \bibinfo
  {author} {\bibfnamefont {D.}~\bibnamefont {Driebe}},\ }\href {\doibase
  10.1016/J.PHYSLETA.2009.12.042} {\bibfield  {journal} {\bibinfo  {journal}
  {Phys. Lett. A}\ }\textbf {\bibinfo {volume} {374}},\ \bibinfo {pages} {1001}
  (\bibinfo {year} {2010})}\BibitemShut {NoStop}%
\bibitem [{\citenamefont {Niedenzu}\ \emph {et~al.}(2018)\citenamefont
  {Niedenzu}, \citenamefont {Mukherjee}, \citenamefont {Ghosh}, \citenamefont
  {Kofman},\ and\ \citenamefont {Kurizki}}]{Niedenzu2018}%
  \BibitemOpen
  \bibfield  {author} {\bibinfo {author} {\bibfnamefont {W.}~\bibnamefont
  {Niedenzu}}, \bibinfo {author} {\bibfnamefont {V.}~\bibnamefont {Mukherjee}},
  \bibinfo {author} {\bibfnamefont {A.}~\bibnamefont {Ghosh}}, \bibinfo
  {author} {\bibfnamefont {A.~G.}\ \bibnamefont {Kofman}}, \ and\ \bibinfo
  {author} {\bibfnamefont {G.}~\bibnamefont {Kurizki}},\ }\href {\doibase
  10.1038/s41467-017-01991-6} {\bibfield  {journal} {\bibinfo  {journal} {Nat.
  Comms.}\ }\textbf {\bibinfo {volume} {9}},\ \bibinfo {pages} {165} (\bibinfo
  {year} {2018})}\BibitemShut {NoStop}%
\bibitem [{\citenamefont {Oppenheim}\ \emph {et~al.}(2002)\citenamefont
  {Oppenheim}, \citenamefont {Horodecki}, \citenamefont {Horodecki},\ and\
  \citenamefont {Horodecki}}]{Oppenheim2002}%
  \BibitemOpen
  \bibfield  {author} {\bibinfo {author} {\bibfnamefont {J.}~\bibnamefont
  {Oppenheim}}, \bibinfo {author} {\bibfnamefont {M.}~\bibnamefont
  {Horodecki}}, \bibinfo {author} {\bibfnamefont {P.}~\bibnamefont
  {Horodecki}}, \ and\ \bibinfo {author} {\bibfnamefont {R.}~\bibnamefont
  {Horodecki}},\ }\href {\doibase 10.1103/PhysRevLett.89.180402} {\bibfield
  {journal} {\bibinfo  {journal} {Phys. Rev. Lett.}\ }\textbf {\bibinfo
  {volume} {89}},\ \bibinfo {pages} {180402} (\bibinfo {year}
  {2002})}\BibitemShut {NoStop}%
\bibitem [{\citenamefont {Perarnau-Llobet}\ \emph {et~al.}(2015)\citenamefont
  {Perarnau-Llobet}, \citenamefont {Hovhannisyan}, \citenamefont {Huber},
  \citenamefont {Skrzypczyk}, \citenamefont {Brunner},\ and\ \citenamefont
  {Ac{\'{i}}n}}]{Perarnau-Llobet2015}%
  \BibitemOpen
  \bibfield  {author} {\bibinfo {author} {\bibfnamefont {M.}~\bibnamefont
  {Perarnau-Llobet}}, \bibinfo {author} {\bibfnamefont {K.~V.}\ \bibnamefont
  {Hovhannisyan}}, \bibinfo {author} {\bibfnamefont {M.}~\bibnamefont {Huber}},
  \bibinfo {author} {\bibfnamefont {P.}~\bibnamefont {Skrzypczyk}}, \bibinfo
  {author} {\bibfnamefont {N.}~\bibnamefont {Brunner}}, \ and\ \bibinfo
  {author} {\bibfnamefont {A.}~\bibnamefont {Ac{\'{i}}n}},\ }\href {\doibase
  10.1103/PhysRevX.5.041011} {\bibfield  {journal} {\bibinfo  {journal} {Phys.
  Rev. X}\ }\textbf {\bibinfo {volume} {5}},\ \bibinfo {pages} {041011}
  (\bibinfo {year} {2015})}\BibitemShut {NoStop}%
\bibitem [{\citenamefont {Zurek}(2003)}]{Zurek2003}%
  \BibitemOpen
  \bibfield  {author} {\bibinfo {author} {\bibfnamefont {W.~H.}\ \bibnamefont
  {Zurek}},\ }\href {\doibase 10.1103/PhysRevA.67.012320} {\bibfield  {journal}
  {\bibinfo  {journal} {Phys. Rev. A}\ }\textbf {\bibinfo {volume} {67}},\
  \bibinfo {pages} {012320} (\bibinfo {year} {2003})}\BibitemShut {NoStop}%
\bibitem [{\citenamefont {Jevtic}\ \emph {et~al.}(2012)\citenamefont {Jevtic},
  \citenamefont {Jennings},\ and\ \citenamefont {Rudolph}}]{Jevtic2012}%
  \BibitemOpen
  \bibfield  {author} {\bibinfo {author} {\bibfnamefont {S.}~\bibnamefont
  {Jevtic}}, \bibinfo {author} {\bibfnamefont {D.}~\bibnamefont {Jennings}}, \
  and\ \bibinfo {author} {\bibfnamefont {T.}~\bibnamefont {Rudolph}},\ }\href
  {\doibase 10.1103/PhysRevLett.108.110403} {\bibfield  {journal} {\bibinfo
  {journal} {Phys. Rev. Lett.}\ }\textbf {\bibinfo {volume} {108}},\ \bibinfo
  {pages} {110403} (\bibinfo {year} {2012})}\BibitemShut {NoStop}%
\bibitem [{\citenamefont {del Rio}\ \emph {et~al.}(2011)\citenamefont {del
  Rio}, \citenamefont {{\AA}berg}, \citenamefont {Renner}, \citenamefont
  {Dahlsten},\ and\ \citenamefont {Vedral}}]{Rio2011}%
  \BibitemOpen
  \bibfield  {author} {\bibinfo {author} {\bibfnamefont {L.}~\bibnamefont {del
  Rio}}, \bibinfo {author} {\bibfnamefont {J.}~\bibnamefont {{\AA}berg}},
  \bibinfo {author} {\bibfnamefont {R.}~\bibnamefont {Renner}}, \bibinfo
  {author} {\bibfnamefont {O.}~\bibnamefont {Dahlsten}}, \ and\ \bibinfo
  {author} {\bibfnamefont {V.}~\bibnamefont {Vedral}},\ }\href {\doibase
  10.1038/nature10123} {\bibfield  {journal} {\bibinfo  {journal} {Nat.}\
  }\textbf {\bibinfo {volume} {474}},\ \bibinfo {pages} {61} (\bibinfo {year}
  {2011})}\BibitemShut {NoStop}%
\bibitem [{\citenamefont {Esposito}\ \emph {et~al.}(2010)\citenamefont
  {Esposito}, \citenamefont {Lindenberg},\ and\ \citenamefont {{Van den
  Broeck}}}]{Esposito2010}%
  \BibitemOpen
  \bibfield  {author} {\bibinfo {author} {\bibfnamefont {M.}~\bibnamefont
  {Esposito}}, \bibinfo {author} {\bibfnamefont {K.}~\bibnamefont
  {Lindenberg}}, \ and\ \bibinfo {author} {\bibfnamefont {C.}~\bibnamefont
  {{Van den Broeck}}},\ }\href {\doibase 10.1088/1367-2630/12/1/013013}
  {\bibfield  {journal} {\bibinfo  {journal} {New J. Phys.}\ }\textbf {\bibinfo
  {volume} {12}},\ \bibinfo {pages} {013013} (\bibinfo {year}
  {2010})}\BibitemShut {NoStop}%
\bibitem [{\citenamefont {Lostaglio}\ \emph
  {et~al.}(2015{\natexlab{a}})\citenamefont {Lostaglio}, \citenamefont
  {Jennings},\ and\ \citenamefont {Rudolph}}]{Lostaglio2015}%
  \BibitemOpen
  \bibfield  {author} {\bibinfo {author} {\bibfnamefont {M.}~\bibnamefont
  {Lostaglio}}, \bibinfo {author} {\bibfnamefont {D.}~\bibnamefont {Jennings}},
  \ and\ \bibinfo {author} {\bibfnamefont {T.}~\bibnamefont {Rudolph}},\ }\href
  {\doibase 10.1038/ncomms7383} {\bibfield  {journal} {\bibinfo  {journal}
  {Nat. Comms.}\ }\textbf {\bibinfo {volume} {6}},\ \bibinfo {pages} {6383}
  (\bibinfo {year} {2015}{\natexlab{a}})}\BibitemShut {NoStop}%
\bibitem [{\citenamefont {Korzekwa}\ \emph {et~al.}(2016)\citenamefont
  {Korzekwa}, \citenamefont {Lostaglio}, \citenamefont {Oppenheim},\ and\
  \citenamefont {Jennings}}]{Korzekwa2016}%
  \BibitemOpen
  \bibfield  {author} {\bibinfo {author} {\bibfnamefont {K.}~\bibnamefont
  {Korzekwa}}, \bibinfo {author} {\bibfnamefont {M.}~\bibnamefont {Lostaglio}},
  \bibinfo {author} {\bibfnamefont {J.}~\bibnamefont {Oppenheim}}, \ and\
  \bibinfo {author} {\bibfnamefont {D.}~\bibnamefont {Jennings}},\ }\href
  {\doibase 10.1088/1367-2630/18/2/023045} {\bibfield  {journal} {\bibinfo
  {journal} {New J. Phys.}\ }\textbf {\bibinfo {volume} {18}},\ \bibinfo
  {pages} {023045} (\bibinfo {year} {2016})}\BibitemShut {NoStop}%
\bibitem [{\citenamefont {{\AA}berg}(2014)}]{Aberg2014}%
  \BibitemOpen
  \bibfield  {author} {\bibinfo {author} {\bibfnamefont {J.}~\bibnamefont
  {{\AA}berg}},\ }\href {\doibase 10.1103/PhysRevLett.113.150402} {\bibfield
  {journal} {\bibinfo  {journal} {Phys. Rev. Lett.}\ }\textbf {\bibinfo
  {volume} {113}},\ \bibinfo {pages} {150402} (\bibinfo {year}
  {2014})}\BibitemShut {NoStop}%
\bibitem [{\citenamefont {Baumgratz}\ \emph {et~al.}(2014)\citenamefont
  {Baumgratz}, \citenamefont {Cramer},\ and\ \citenamefont
  {Plenio}}]{Baumgratz2014}%
  \BibitemOpen
  \bibfield  {author} {\bibinfo {author} {\bibfnamefont {T.}~\bibnamefont
  {Baumgratz}}, \bibinfo {author} {\bibfnamefont {M.}~\bibnamefont {Cramer}}, \
  and\ \bibinfo {author} {\bibfnamefont {M.~B.}\ \bibnamefont {Plenio}},\
  }\href {\doibase 10.1103/PhysRevLett.113.140401} {\bibfield  {journal}
  {\bibinfo  {journal} {Phys. Rev. Lett.}\ }\textbf {\bibinfo {volume} {113}},\
  \bibinfo {pages} {140401} (\bibinfo {year} {2014})}\BibitemShut {NoStop}%
\bibitem [{\citenamefont {Lostaglio}\ \emph
  {et~al.}(2015{\natexlab{b}})\citenamefont {Lostaglio}, \citenamefont
  {Korzekwa}, \citenamefont {Jennings},\ and\ \citenamefont
  {Rudolph}}]{Lostaglio2015X}%
  \BibitemOpen
  \bibfield  {author} {\bibinfo {author} {\bibfnamefont {M.}~\bibnamefont
  {Lostaglio}}, \bibinfo {author} {\bibfnamefont {K.}~\bibnamefont {Korzekwa}},
  \bibinfo {author} {\bibfnamefont {D.}~\bibnamefont {Jennings}}, \ and\
  \bibinfo {author} {\bibfnamefont {T.}~\bibnamefont {Rudolph}},\ }\href
  {\doibase 10.1103/PhysRevX.5.021001} {\bibfield  {journal} {\bibinfo
  {journal} {Phys. Rev. X}\ }\textbf {\bibinfo {volume} {5}},\ \bibinfo {pages}
  {021001} (\bibinfo {year} {2015}{\natexlab{b}})}\BibitemShut {NoStop}%
\bibitem [{\citenamefont {Marvian}\ \emph {et~al.}(2016)\citenamefont
  {Marvian}, \citenamefont {Spekkens},\ and\ \citenamefont
  {Zanardi}}]{Marvian2016}%
  \BibitemOpen
  \bibfield  {author} {\bibinfo {author} {\bibfnamefont {I.}~\bibnamefont
  {Marvian}}, \bibinfo {author} {\bibfnamefont {R.~W.}\ \bibnamefont
  {Spekkens}}, \ and\ \bibinfo {author} {\bibfnamefont {P.}~\bibnamefont
  {Zanardi}},\ }\href {\doibase 10.1103/PhysRevA.93.052331} {\bibfield
  {journal} {\bibinfo  {journal} {Phys. Rev. A}\ }\textbf {\bibinfo {volume}
  {93}},\ \bibinfo {pages} {052331} (\bibinfo {year} {2016})}\BibitemShut
  {NoStop}%
\bibitem [{\citenamefont {Scully}\ \emph {et~al.}(2003)\citenamefont {Scully},
  \citenamefont {Zubairy}, \citenamefont {Agarwal},\ and\ \citenamefont
  {Walther}}]{2003scully862}%
  \BibitemOpen
  \bibfield  {author} {\bibinfo {author} {\bibfnamefont {M.~O.}\ \bibnamefont
  {Scully}}, \bibinfo {author} {\bibfnamefont {M.~S.}\ \bibnamefont {Zubairy}},
  \bibinfo {author} {\bibfnamefont {G.~S.}\ \bibnamefont {Agarwal}}, \ and\
  \bibinfo {author} {\bibfnamefont {H.}~\bibnamefont {Walther}},\ }\href
  {\doibase 10.1126/science.1078955} {\bibfield  {journal} {\bibinfo  {journal}
  {Sci.}\ }\textbf {\bibinfo {volume} {299}},\ \bibinfo {pages} {862} (\bibinfo
  {year} {2003})}\BibitemShut {NoStop}%
\bibitem [{\citenamefont {Quan}\ \emph {et~al.}(2006)\citenamefont {Quan},
  \citenamefont {Zhang},\ and\ \citenamefont {Sun}}]{Quan2006}%
  \BibitemOpen
  \bibfield  {author} {\bibinfo {author} {\bibfnamefont {H.~T.}\ \bibnamefont
  {Quan}}, \bibinfo {author} {\bibfnamefont {P.}~\bibnamefont {Zhang}}, \ and\
  \bibinfo {author} {\bibfnamefont {C.~P.}\ \bibnamefont {Sun}},\ }\href
  {\doibase 10.1103/PhysRevE.73.036122} {\bibfield  {journal} {\bibinfo
  {journal} {Phys. Rev. E}\ }\textbf {\bibinfo {volume} {73}},\ \bibinfo
  {pages} {036122} (\bibinfo {year} {2006})}\BibitemShut {NoStop}%
\bibitem [{\citenamefont {Dillenschneider}\ and\ \citenamefont
  {Lutz}(2009)}]{Dillenschneider2009}%
  \BibitemOpen
  \bibfield  {author} {\bibinfo {author} {\bibfnamefont {R.}~\bibnamefont
  {Dillenschneider}}\ and\ \bibinfo {author} {\bibfnamefont {E.}~\bibnamefont
  {Lutz}},\ }\href {\doibase 10.1209/0295-5075/88/50003} {\bibfield  {journal}
  {\bibinfo  {journal} {EPL}\ }\textbf {\bibinfo {volume} {88}},\ \bibinfo
  {pages} {50003} (\bibinfo {year} {2009})}\BibitemShut {NoStop}%
\bibitem [{\citenamefont {T{\"{u}}rkpen{\c{c}}e}\ \emph
  {et~al.}(2017)\citenamefont {T{\"{u}}rkpen{\c{c}}e}, \citenamefont
  {Altintas}, \citenamefont {Paternostro},\ and\ \citenamefont
  {M{\"{u}}stecapl{\i}o{\u{g}}lu}}]{Turkpence2017}%
  \BibitemOpen
  \bibfield  {author} {\bibinfo {author} {\bibfnamefont {D.}~\bibnamefont
  {T{\"{u}}rkpen{\c{c}}e}}, \bibinfo {author} {\bibfnamefont {F.}~\bibnamefont
  {Altintas}}, \bibinfo {author} {\bibfnamefont {M.}~\bibnamefont
  {Paternostro}}, \ and\ \bibinfo {author} {\bibfnamefont {{\"{O}}.~E.}\
  \bibnamefont {M{\"{u}}stecapl{\i}o{\u{g}}lu}},\ }\href {\doibase
  10.1209/0295-5075/117/50002} {\bibfield  {journal} {\bibinfo  {journal}
  {EPL}\ }\textbf {\bibinfo {volume} {117}},\ \bibinfo {pages} {50002}
  (\bibinfo {year} {2017})}\BibitemShut {NoStop}%
\bibitem [{\citenamefont {Da{\u{g}}}\ \emph {et~al.}(2016)\citenamefont
  {Da{\u{g}}}, \citenamefont {Niedenzu}, \citenamefont
  {M{\"{u}}stecapl{\i}o{\u{g}}lu},\ and\ \citenamefont {Kurizki}}]{Dag2016}%
  \BibitemOpen
  \bibfield  {author} {\bibinfo {author} {\bibfnamefont {C.~B.}\ \bibnamefont
  {Da{\u{g}}}}, \bibinfo {author} {\bibfnamefont {W.}~\bibnamefont {Niedenzu}},
  \bibinfo {author} {\bibfnamefont {{\"{O}}.~E.}\ \bibnamefont
  {M{\"{u}}stecapl{\i}o{\u{g}}lu}}, \ and\ \bibinfo {author} {\bibfnamefont
  {G.}~\bibnamefont {Kurizki}},\ }\href {\doibase 10.3390/e18070244} {\bibfield
   {journal} {\bibinfo  {journal} {Ent.}\ }\textbf {\bibinfo {volume} {18}},\
  \bibinfo {pages} {244} (\bibinfo {year} {2016})}\BibitemShut {NoStop}%
\bibitem [{\citenamefont {Da{\u{g}}}\ \emph {et~al.}(2019)\citenamefont
  {Da{\u{g}}}, \citenamefont {Niedenzu}, \citenamefont {Ozaydin}, \citenamefont
  {M{\"{u}}stecapl{\i}o{\u{g}}lu},\ and\ \citenamefont {Kurizki}}]{Dag2018}%
  \BibitemOpen
  \bibfield  {author} {\bibinfo {author} {\bibfnamefont {C.~B.}\ \bibnamefont
  {Da{\u{g}}}}, \bibinfo {author} {\bibfnamefont {W.}~\bibnamefont {Niedenzu}},
  \bibinfo {author} {\bibfnamefont {F.}~\bibnamefont {Ozaydin}}, \bibinfo
  {author} {\bibfnamefont {{\"{O}}.~E.}\ \bibnamefont
  {M{\"{u}}stecapl{\i}o{\u{g}}lu}}, \ and\ \bibinfo {author} {\bibfnamefont
  {G.}~\bibnamefont {Kurizki}},\ }\href {\doibase 10.1021/acs.jpcc.8b11445}
  {\bibfield  {journal} {\bibinfo  {journal} {J. Phys. Chem. C}\ }\textbf
  {\bibinfo {volume} {123}},\ \bibinfo {pages} {4035} (\bibinfo {year}
  {2019})}\BibitemShut {NoStop}%
\bibitem [{\citenamefont {Hardal}\ and\ \citenamefont
  {M{\"{u}}stecapl{\i}o{\u{g}}lu}(2015)}]{Hardal2015}%
  \BibitemOpen
  \bibfield  {author} {\bibinfo {author} {\bibfnamefont {A.~{\"{U}}.~C.}\
  \bibnamefont {Hardal}}\ and\ \bibinfo {author} {\bibfnamefont {{\"{O}}.~E.}\
  \bibnamefont {M{\"{u}}stecapl{\i}o{\u{g}}lu}},\ }\href {\doibase
  10.1038/srep12953} {\bibfield  {journal} {\bibinfo  {journal} {Sci. Rep.}\
  }\textbf {\bibinfo {volume} {5}},\ \bibinfo {pages} {12953} (\bibinfo {year}
  {2015})}\BibitemShut {NoStop}%
\bibitem [{\citenamefont {Curzon}\ and\ \citenamefont
  {Ahlborn}(1975)}]{Curzon1975}%
  \BibitemOpen
  \bibfield  {author} {\bibinfo {author} {\bibfnamefont {F.~L.}\ \bibnamefont
  {Curzon}}\ and\ \bibinfo {author} {\bibfnamefont {B.}~\bibnamefont
  {Ahlborn}},\ }\href {\doibase 10.1119/1.10023} {\bibfield  {journal}
  {\bibinfo  {journal} {Am. J. Phys.}\ }\textbf {\bibinfo {volume} {43}},\
  \bibinfo {pages} {22} (\bibinfo {year} {1975})}\BibitemShut {NoStop}%
\bibitem [{\citenamefont {Torrontegui}\ \emph {et~al.}(2013)\citenamefont
  {Torrontegui} \emph {et~al.}}]{ATOMTorrontegui2013}%
  \BibitemOpen
  \bibfield  {author} {\bibinfo {author} {\bibfnamefont {E.}~\bibnamefont
  {Torrontegui}} \emph {et~al.},\ }\href {\doibase
  10.1016/B978-0-12-408090-4.00002-5} {\bibfield  {journal} {\bibinfo
  {journal} {Adv. Atom. Mol. Opt. Phys.}\ }\textbf {\bibinfo {volume} {62}},\
  \bibinfo {pages} {107} (\bibinfo {year} {2013})}\BibitemShut {NoStop}%
\bibitem [{\citenamefont {del Campo}\ \emph {et~al.}(2014)\citenamefont {del
  Campo}, \citenamefont {Goold},\ and\ \citenamefont
  {Paternostro}}]{Campo2015}%
  \BibitemOpen
  \bibfield  {author} {\bibinfo {author} {\bibfnamefont {A.}~\bibnamefont {del
  Campo}}, \bibinfo {author} {\bibfnamefont {J.}~\bibnamefont {Goold}}, \ and\
  \bibinfo {author} {\bibfnamefont {M.}~\bibnamefont {Paternostro}},\ }\href
  {\doibase 10.1038/srep06208} {\bibfield  {journal} {\bibinfo  {journal} {Sci.
  Rep.}\ }\textbf {\bibinfo {volume} {4}},\ \bibinfo {pages} {6208} (\bibinfo
  {year} {2014})}\BibitemShut {NoStop}%
\bibitem [{\citenamefont {Beau}\ \emph {et~al.}(2016)\citenamefont {Beau},
  \citenamefont {Jaramillo}, \citenamefont {del Campo}, \citenamefont {Beau},
  \citenamefont {Jaramillo},\ and\ \citenamefont {del Campo}}]{Beau2016}%
  \BibitemOpen
  \bibfield  {author} {\bibinfo {author} {\bibfnamefont {M.}~\bibnamefont
  {Beau}}, \bibinfo {author} {\bibfnamefont {J.}~\bibnamefont {Jaramillo}},
  \bibinfo {author} {\bibfnamefont {A.}~\bibnamefont {del Campo}}, \bibinfo
  {author} {\bibfnamefont {M.}~\bibnamefont {Beau}}, \bibinfo {author}
  {\bibfnamefont {J.}~\bibnamefont {Jaramillo}}, \ and\ \bibinfo {author}
  {\bibfnamefont {A.}~\bibnamefont {del Campo}},\ }\href {\doibase
  10.3390/e18050168} {\bibfield  {journal} {\bibinfo  {journal} {Ent.}\
  }\textbf {\bibinfo {volume} {18}},\ \bibinfo {pages} {168} (\bibinfo {year}
  {2016})}\BibitemShut {NoStop}%
\bibitem [{\citenamefont {Abah}\ and\ \citenamefont
  {Paternostro}(2019)}]{Abah2018}%
  \BibitemOpen
  \bibfield  {author} {\bibinfo {author} {\bibfnamefont {O.}~\bibnamefont
  {Abah}}\ and\ \bibinfo {author} {\bibfnamefont {M.}~\bibnamefont
  {Paternostro}},\ }\href {\doibase 10.1103/PhysRevE.99.022110} {\bibfield
  {journal} {\bibinfo  {journal} {Phys. Rev. E}\ }\textbf {\bibinfo {volume}
  {99}},\ \bibinfo {pages} {022110} (\bibinfo {year} {2019})}\BibitemShut
  {NoStop}%
\bibitem [{\citenamefont {Abah}\ and\ \citenamefont
  {Lutz}(2018)}]{Abah2018PRE}%
  \BibitemOpen
  \bibfield  {author} {\bibinfo {author} {\bibfnamefont {O.}~\bibnamefont
  {Abah}}\ and\ \bibinfo {author} {\bibfnamefont {E.}~\bibnamefont {Lutz}},\
  }\href {\doibase 10.1103/PhysRevE.98.032121} {\bibfield  {journal} {\bibinfo
  {journal} {Phys. Rev. E}\ }\textbf {\bibinfo {volume} {98}},\ \bibinfo
  {pages} {032121} (\bibinfo {year} {2018})}\BibitemShut {NoStop}%
\bibitem [{\citenamefont {Abah}\ and\ \citenamefont
  {Lutz}(2017)}]{EPLLutz2017}%
  \BibitemOpen
  \bibfield  {author} {\bibinfo {author} {\bibfnamefont {O.}~\bibnamefont
  {Abah}}\ and\ \bibinfo {author} {\bibfnamefont {E.}~\bibnamefont {Lutz}},\
  }\href {\doibase 10.1209/0295-5075/118/40005} {\bibfield  {journal} {\bibinfo
   {journal} {EPL}\ }\textbf {\bibinfo {volume} {118}},\ \bibinfo {pages}
  {40005} (\bibinfo {year} {2017})}\BibitemShut {NoStop}%
\bibitem [{\citenamefont {T{\"{u}}rkpen{\c{c}}e}\ and\ \citenamefont
  {M{\"{u}}stecapl{\i}o{\u{g}}lu}(2016)}]{Turkpence2016}%
  \BibitemOpen
  \bibfield  {author} {\bibinfo {author} {\bibfnamefont {D.}~\bibnamefont
  {T{\"{u}}rkpen{\c{c}}e}}\ and\ \bibinfo {author} {\bibfnamefont
  {{\"{O}}.~E.}\ \bibnamefont {M{\"{u}}stecapl{\i}o{\u{g}}lu}},\ }\href
  {\doibase 10.1103/PhysRevE.93.012145} {\bibfield  {journal} {\bibinfo
  {journal} {Phys. Rev. E}\ }\textbf {\bibinfo {volume} {93}},\ \bibinfo
  {pages} {012145} (\bibinfo {year} {2016})}\BibitemShut {NoStop}%
\bibitem [{\citenamefont {Deffner}\ \emph {et~al.}(2010)\citenamefont
  {Deffner}, \citenamefont {Abah},\ and\ \citenamefont {Lutz}}]{CHEMDeff2010}%
  \BibitemOpen
  \bibfield  {author} {\bibinfo {author} {\bibfnamefont {S.}~\bibnamefont
  {Deffner}}, \bibinfo {author} {\bibfnamefont {O.}~\bibnamefont {Abah}}, \
  and\ \bibinfo {author} {\bibfnamefont {E.}~\bibnamefont {Lutz}},\ }\href
  {\doibase 10.1016/J.CHEMPHYS.2010.04.042} {\bibfield  {journal} {\bibinfo
  {journal} {Chem. Phys.}\ }\textbf {\bibinfo {volume} {375}},\ \bibinfo
  {pages} {200} (\bibinfo {year} {2010})}\BibitemShut {NoStop}%
\bibitem [{\citenamefont {Husimi}(1953)}]{PTPHusimi1953}%
  \BibitemOpen
  \bibfield  {author} {\bibinfo {author} {\bibfnamefont {K.}~\bibnamefont
  {Husimi}},\ }\href {\doibase 10.1143/ptp/9.4.381} {\bibfield  {journal}
  {\bibinfo  {journal} {Prog. T. Phys.}\ }\textbf {\bibinfo {volume} {9}},\
  \bibinfo {pages} {381} (\bibinfo {year} {1953})}\BibitemShut {NoStop}%
\bibitem [{\citenamefont {Deng}\ \emph {et~al.}(2018)\citenamefont {Deng},
  \citenamefont {Chenu}, \citenamefont {Diao}, \citenamefont {Li},
  \citenamefont {Yu}, \citenamefont {Coulamy}, \citenamefont {del Campo},\ and\
  \citenamefont {Wu}}]{Deng2018}%
  \BibitemOpen
  \bibfield  {author} {\bibinfo {author} {\bibfnamefont {S.}~\bibnamefont
  {Deng}}, \bibinfo {author} {\bibfnamefont {A.}~\bibnamefont {Chenu}},
  \bibinfo {author} {\bibfnamefont {P.}~\bibnamefont {Diao}}, \bibinfo {author}
  {\bibfnamefont {F.}~\bibnamefont {Li}}, \bibinfo {author} {\bibfnamefont
  {S.}~\bibnamefont {Yu}}, \bibinfo {author} {\bibfnamefont {I.}~\bibnamefont
  {Coulamy}}, \bibinfo {author} {\bibfnamefont {A.}~\bibnamefont {del Campo}},
  \ and\ \bibinfo {author} {\bibfnamefont {H.}~\bibnamefont {Wu}},\ }\href
  {\doibase 10.1126/sciadv.aar5909} {\bibfield  {journal} {\bibinfo  {journal}
  {Sci. Adv.}\ }\textbf {\bibinfo {volume} {4}},\ \bibinfo {pages} {eaar5909}
  (\bibinfo {year} {2018})}\BibitemShut {NoStop}%
\bibitem [{\citenamefont {Diao}\ \emph {et~al.}(2018)\citenamefont {Diao},
  \citenamefont {Deng}, \citenamefont {Li}, \citenamefont {Yu}, \citenamefont
  {Chenu}, \citenamefont {del Campo},\ and\ \citenamefont {Wu}}]{Diao2018}%
  \BibitemOpen
  \bibfield  {author} {\bibinfo {author} {\bibfnamefont {P.}~\bibnamefont
  {Diao}}, \bibinfo {author} {\bibfnamefont {S.}~\bibnamefont {Deng}}, \bibinfo
  {author} {\bibfnamefont {F.}~\bibnamefont {Li}}, \bibinfo {author}
  {\bibfnamefont {S.}~\bibnamefont {Yu}}, \bibinfo {author} {\bibfnamefont
  {A.}~\bibnamefont {Chenu}}, \bibinfo {author} {\bibfnamefont
  {A.}~\bibnamefont {del Campo}}, \ and\ \bibinfo {author} {\bibfnamefont
  {H.}~\bibnamefont {Wu}},\ }\href {\doibase 10.1088/1367-2630/aae45e}
  {\bibfield  {journal} {\bibinfo  {journal} {New J. Phys.}\ }\textbf {\bibinfo
  {volume} {20}},\ \bibinfo {pages} {105004} (\bibinfo {year}
  {2018})}\BibitemShut {NoStop}%
\bibitem [{\citenamefont {del Campo}(2013)}]{DelCampo2013}%
  \BibitemOpen
  \bibfield  {author} {\bibinfo {author} {\bibfnamefont {A.}~\bibnamefont {del
  Campo}},\ }\href {\doibase 10.1103/PhysRevLett.111.100502} {\bibfield
  {journal} {\bibinfo  {journal} {Phys. Rev. Lett.}\ }\textbf {\bibinfo
  {volume} {111}},\ \bibinfo {pages} {100502} (\bibinfo {year}
  {2013})}\BibitemShut {NoStop}%
\bibitem [{\citenamefont {Kosloff}\ and\ \citenamefont
  {Rezek}(2017)}]{Kosloff2017}%
  \BibitemOpen
  \bibfield  {author} {\bibinfo {author} {\bibfnamefont {R.}~\bibnamefont
  {Kosloff}}\ and\ \bibinfo {author} {\bibfnamefont {Y.}~\bibnamefont
  {Rezek}},\ }\href {\doibase 10.3390/e19040136} {\bibfield  {journal}
  {\bibinfo  {journal} {Entropy}\ }\textbf {\bibinfo {volume} {19}},\ \bibinfo
  {pages} {136} (\bibinfo {year} {2017})}\BibitemShut {NoStop}%
\bibitem [{\citenamefont {Torrontegui}\ \emph {et~al.}(2017)\citenamefont
  {Torrontegui}, \citenamefont {Lizuain}, \citenamefont
  {Gonz{\'{a}}lez-Resines}, \citenamefont {Tobalina}, \citenamefont
  {Ruschhaupt}, \citenamefont {Kosloff},\ and\ \citenamefont
  {Muga}}]{Torrontegui2017}%
  \BibitemOpen
  \bibfield  {author} {\bibinfo {author} {\bibfnamefont {E.}~\bibnamefont
  {Torrontegui}}, \bibinfo {author} {\bibfnamefont {I.}~\bibnamefont
  {Lizuain}}, \bibinfo {author} {\bibfnamefont {S.}~\bibnamefont
  {Gonz{\'{a}}lez-Resines}}, \bibinfo {author} {\bibfnamefont {A.}~\bibnamefont
  {Tobalina}}, \bibinfo {author} {\bibfnamefont {A.}~\bibnamefont
  {Ruschhaupt}}, \bibinfo {author} {\bibfnamefont {R.}~\bibnamefont {Kosloff}},
  \ and\ \bibinfo {author} {\bibfnamefont {J.~G.}\ \bibnamefont {Muga}},\
  }\href {\doibase 10.1103/PhysRevA.96.022133} {\bibfield  {journal} {\bibinfo
  {journal} {Physical Review A}\ }\textbf {\bibinfo {volume} {96}},\ \bibinfo
  {pages} {022133} (\bibinfo {year} {2017})}\BibitemShut {NoStop}%
\bibitem [{\citenamefont {Zheng}\ \emph {et~al.}(2016)\citenamefont {Zheng},
  \citenamefont {Campbell}, \citenamefont {{De Chiara}},\ and\ \citenamefont
  {Poletti}}]{Zheng2016}%
  \BibitemOpen
  \bibfield  {author} {\bibinfo {author} {\bibfnamefont {Y.}~\bibnamefont
  {Zheng}}, \bibinfo {author} {\bibfnamefont {S.}~\bibnamefont {Campbell}},
  \bibinfo {author} {\bibfnamefont {G.}~\bibnamefont {{De Chiara}}}, \ and\
  \bibinfo {author} {\bibfnamefont {D.}~\bibnamefont {Poletti}},\ }\href
  {\doibase 10.1103/PhysRevA.94.042132} {\bibfield  {journal} {\bibinfo
  {journal} {Physical Review A}\ }\textbf {\bibinfo {volume} {94}},\ \bibinfo
  {pages} {042132} (\bibinfo {year} {2016})}\BibitemShut {NoStop}%
\bibitem [{\citenamefont {Calzetta}(2018)}]{Calzetta2018}%
  \BibitemOpen
  \bibfield  {author} {\bibinfo {author} {\bibfnamefont {E.}~\bibnamefont
  {Calzetta}},\ }\href {\doibase 10.1103/PhysRevA.98.032107} {\bibfield
  {journal} {\bibinfo  {journal} {Physical Review A}\ }\textbf {\bibinfo
  {volume} {98}},\ \bibinfo {pages} {032107} (\bibinfo {year}
  {2018})}\BibitemShut {NoStop}%
\bibitem [{\citenamefont {Feldmann}\ \emph {et~al.}(1996)\citenamefont
  {Feldmann}, \citenamefont {Geva}, \citenamefont {Kosloff},\ and\
  \citenamefont {Salamon}}]{SCFeldmann1996}%
  \BibitemOpen
  \bibfield  {author} {\bibinfo {author} {\bibfnamefont {T.}~\bibnamefont
  {Feldmann}}, \bibinfo {author} {\bibfnamefont {E.}~\bibnamefont {Geva}},
  \bibinfo {author} {\bibfnamefont {R.}~\bibnamefont {Kosloff}}, \ and\
  \bibinfo {author} {\bibfnamefont {P.}~\bibnamefont {Salamon}},\ }\href
  {\doibase 10.1119/1.18197} {\bibfield  {journal} {\bibinfo  {journal} {Am. J.
  Phys.}\ }\textbf {\bibinfo {volume} {64}},\ \bibinfo {pages} {485} (\bibinfo
  {year} {1996})}\BibitemShut {NoStop}%
\end{thebibliography}%

\end{document}